\documentclass[reprint,amsmath,amssymb, aps,,superscriptaddress,notitlepage,nofootinbib]{revtex4-1}

\usepackage{graphicx}
\usepackage{dcolumn}
\usepackage{bm}
\usepackage[dvipsnames]{xcolor}
\usepackage{hyperref}

\newcommand{\GRAPPA}{GRAPPA Institute, University of Amsterdam, 1098 XH Amsterdam, The Netherlands}
\newcommand{\UvA}{Institute for Theoretical Physics, University of Amsterdam, 1098 XH Amsterdam, The Netherlands}
\newcommand{\KAVLI}{Kavli Institute for the Physics and Mathematics of the Universe (Kavli IPMU, WPI), University of Tokyo, Kashiwa, Chiba 277-8583, Japan}

\begin{document}

\title{Searches for sterile neutrinos and axionlike particles from the Galactic halo with eROSITA}

\author{Ariane Dekker}
\email{a.h.dekker@uva.nl}
\thanks{\scriptsize \!\!
\href{https://orcid.org/0000-0002-3831-9442}{orcid.org/0000-0002-3831-9442}}
\affiliation{\GRAPPA}
\affiliation{\UvA}

\author{Ebo Peerbooms}
\email{e.peerbooms@uva.nl}
\affiliation{\UvA}

\author{Fabian Zimmer}
\email{fabian.zimmer@student.uva.nl}
\thanks{\scriptsize \!\!
\href{https://orcid.org/0000-0001-7574-9313}{orcid.org/0000-0001-7574-931}}
\affiliation{\GRAPPA}

\author{Kenny C. Y. Ng}
\email{kcyng@cuhk.edu.hk}
\thanks{\scriptsize \!\! \href{http://orcid.org/0000-0001-8016-2170}{orcid.org/0000-0001-8016-2170}}
\affiliation{Department of Physics, The Chinese University of Hong Kong, Shatin, Hong Kong China}
\affiliation{\GRAPPA}
\affiliation{\UvA}

\author{Shin'ichiro Ando}
\email{s.ando@uva.nl}
\thanks{\scriptsize \!\!
\href{http://orcid.org/0000-0001-6231-7693}{orcid.org/0000-0001-6231-7693}}
\affiliation{\GRAPPA}
\affiliation{\UvA}
\affiliation{\KAVLI}

\date{March 24, 2021}

\begin{abstract}
Dark matter might be made of ``warm'' particles, such as sterile neutrinos in the keV mass range, which can decay into photons through mixing and are consequently detectable by X-ray telescopes. Axionlike particles (ALPs) are detectable by X-ray telescopes too when coupled to standard model particles and decay into photons in the keV range. Both particles could explain the unidentified 3.5-keV line and, interestingly, XENON1T observed an excess of electron recoil events most prominent at $2-3$~keV. One explanation could be an ALPs origin, which is not yet excluded by X-ray constraints in an anomaly-free symmetry model in which the photon production is suppressed. 
We study the diffuse emission coming from the Galactic halo, and calculate the sensitivity of all-sky X-ray survey performed by eROSITA to identify a sterile neutrino or ALP dark matter. We estimate bounds on the mixing angle of the sterile neutrinos and coupling strength of the ALPs. After four years of data-taking by eROSITA, we expect to set stringent constraints, and in particular, we expect to firmly probe mixing angle $\sin^2(2\theta)$ up to nearly two orders magnitude below the best-fit value for explaining the unidentified 3.5-keV line. Moreover, with eROSITA, we will be able to probe the ALP parameter space of couplings to photons and electrons, and potentially confirm an ALP origin of the XENON1T excess.
\end{abstract}

\maketitle

\section{\label{sec:Introduction} Introduction}
The nature of dark matter remains an exciting open question in astrophysics, cosmology and particle physics. The most studied candidate is cold dark matter (CDM), which are particles with negligible thermal velocities during structure formation era and that interact only through gravity with baryons and with themselves~\cite{Frenk_2012}. However, the non-observation of CDM with direct detection experiments motivates to search for other candidates besides CDM; for example, dark matter with much lower masses such as warm dark matter (WDM), with non-negligible thermal velocity at early times~\cite{Roszkowski_2018,Bertone:2010zza}. WDM suppresses structure at small scales due to free-streaming effects, and may explain some CDM issues on structure formation, while it behaves as CDM on large scales~\cite{Bode_2001}.

Sterile neutrinos are well motivated WDM candidates~\cite{PhysRevLett.82.2832,PhysRevD.64.023501,Dolgov:2000ew,Canetti:2012kh}. They have much larger mass than active neutrinos and can explain all of the dark matter density in the Universe~\cite{Boyarsky:2018tvu}. 
Moreover, sterile neutrinos have right-handed chirality in contrary to the three flavour standard model neutrinos which have only left-handed chirality. Presence of right-handed neutrinos is naturally expected in order to explain light neutrino masses found by the neutrino oscillation measurements~\cite{Kajita_1999,Asaka_2005}, and additionally, it may solve the matter-antimatter asymmetry of the Universe~\cite{Asaka_2005}. In the neutrino minimal standard model ($\nu$MSM) it is possible to simultaneous explain baryogenesis, neutrinos mass, and dark matter, with sterile neutrino as the dark matter candidate~\cite{Boyarsky_2009}.

Through a mixing angle $\theta$ with active neutrinos, sterile neutrinos can decay into an active neutrino and a photon with photon energy $E_{\gamma}=m_{\nu_s}/2$, where $m_{\nu_s}$ is the sterile neutrino mass. The rate of decay depends on its mass and mixing angle, and is given by the following decay rate~\cite{Sicilian:2020glg, abazajian2021neutrinos},
\begin{equation}
    \Gamma_{\nu_s}(m_{\nu_s}, \theta) = 1.38 \times 10^{-29} ~ {\rm s}^{-1} \left[ \frac{ \sin^2(2\theta)}{10^{-7}} \right]\left(\frac{m_{\nu_s}}{1~ {\rm keV}} \right)^5 .
\end{equation}
Sterile neutrinos in the keV mass range will produce X-ray photons, which can be observed by X-ray telescopes as a monochromatic line signal.

Through a stacked X-ray spectrum analysis of 73 galaxy clusters, an emission line at $\sim 3.5$~keV was detected~\cite{Bulbul_2014}, hinting towards experimental evidence for sterile neutrino decay.
The same emission line was soon after observed in X-ray spectra of the M31 galaxy and the Perseus galaxy cluster~\cite{Boyarsky_2014}, as well as in the Milky-Way center~\cite{PhysRevLett.115.161301}. The signal can be interpreted as a signature of decaying sterile neutrinos with mass of $m_{\nu_s}\sim 7$~keV and mixing angle of $\sin^2(2\theta) \simeq (0.2$--$2) \times 10^{-10}$~\cite{Boyarsky:2018tvu}. Many follow-up studies confirmed the emission line in spectra of dark matter dominated objects with different X-ray instruments~\cite{iakubovskyi2015testing,Franse_2016,Cappelluti_2018,Neronov_2016}. However, other studies did not detect any line emission in dark matter dominated objects~\cite{urban2014suzaku,Tamura_2015,Riemer_S_rensen_2015,PhysRevD.90.103506,Anderson_2015, foster2021deep,Bhargava_2020} and suggested non-dark matter explanations including astrophysical effects, statistical fluctuations, and instrumental systematics. Recently, using \textit{XMM-Newton} data, Ref.~\cite{Dessert_2020} provides strong constraints on the unassociated X-ray line from decaying dark matter (see also Refs.~\cite{Abazajian:2020unr, Boyarsky:2020hqb}).
It remains  relevant to search for a line emission from sterile neutrino decays. 

Another interesting WDM candidate is the axionlike particle (ALP), which is a pseudo-Nambu-Goldstone boson that emerges when a continuous global symmetry is spontaneously broken~\cite{Arias_2012,Takahashi:2020bpq,Irastorza:2018dyq,Chaubey_2020}. In contrary to the QCD axions, ALPs do not solve the CP problem, and can be light due to the broken symmetry. ALPs can couple to various standard model particles such as protons, electrons, and photons, and therefore can decay into two photons, producing a narrow X-ray line, possibly explaining the unidentified 3.5-keV line.

As another interesting possibility, ALP decay could explain the observed excess of electron recoil events over known backgrounds at the XENON1T experiment, where a best-fit mass value of $m_a=2.3$~keV and coupling to electrons of $g_{ae}\sim10^{-13}$ is found with a $3\sigma$ significance over the background~\cite{Aprile_2020}. The ALP coupling to standard model particles is already too tightly constrained by X-ray observations to explain this excess.
Therefore, the photon coupling needs to be suppressed~\cite{Irastorza:2018dyq}. 
In an anomaly-free symmetry model, the ALP is coupled to leptons without any anomalous coupling to photons, and is dominated by the coupling to the least massive lepton -- the electron~\cite{Nakayama_2014,Pospelov_2008,Takahashi:2020bpq}. In this model, photons are only induced through threshold corrections, and, although suppressed, the decay into two photons can be the leading decay mode for ALPs with masses less than twice the electron mass. 

The ALP-electron coupling suggested by XENON1T is of the same order as the coupling suggested to explain the observed excess in cooling of various stellar objects like white dwarfs and red giants~\cite{Giannotti:2017hny}, known as the stellar cooling anomalies~\cite{Bertolami:2014wua,Ayala:2014pea,Corsico:2012ki,Corsico:2012sh}. The evolution of these objects are described by well-established cooling process, and indeed, the cooling anomaly based on white dwarf luminosity function analysis is found at $4\sigma$~\cite{Giannotti:2015kwo}, hinting towards a preferred region for the ALP parameter space. Following Ref.~\cite{Takahashi:2020bpq}, we consider a model in which ALPs can explain both the XENON1T excess and the stellar cooling anomaly. 

Whether dark matter is made of sterile neutrinos or ALPs, they can be well probed with current generation and future X-ray telescopes. In this paper, we estimate the sensitivity of all-sky X-ray survey performed by eROSITA to observe a decaying sterile neutrino and axion-like particle signal. The hierarchical clustering of dark matter predicts that the Milky-Way galaxy is embedded in a halo of dark matter particles, with a higher density towards the Galactic center~\cite{Navarro_1997}. The largest contribution to the observable dark matter induced X-ray flux originates from the Galactic center, and we study the diffuse emission coming from the Galactic halo around its center. eROSITA has excellent angular and energy resolution, and will also observe the full sky over the course of four years with an average exposure of $2.5$~ks~\cite{merloni2012erosita}, making the survey a valuable probe for dark matter decay with a narrow X-ray line emission. By simulating the all-sky X-ray signal due to dark matter decay, we make a sensitivity projection for eROSITA to a sterile neutrino and ALP signal under a background-only hypothesis. We find that the eROSITA will enable us to probe much deeper regions of the parameter space for both the sterile neutrino and ALP dark matter. 

The paper is organised as follows. In section~\ref{sec:all-sky}, we describe the main features of the all-sky X-ray maps. In section~\ref{sec:anaysis}, we present our analysis methodology, followed by our results and discussions in section~\ref{sec:results}. 
We conclude our paper in Sec.~\ref{sec:conclusion}.

\section{All-sky X-ray map} \label{sec:all-sky}
\subsubsection{Sterile neutrino signal from the Galactic halo}

The X-ray photon flux produced through sterile neutrino decay inside the Galactic halo depends on the sterile neutrino decay rate $\Gamma_{\nu_s}$, sterile neutrino mass $m_{\nu_s}$, energy spectrum $dN_{\rm{decay}}/dE$ per decay, and the dark matter density distribution through the so-called D-factor. The X-ray photon intensity is given as follows,
\begin{equation}
    \frac{d\Phi}{dE} = \frac{\Gamma_{\nu_s}}{4 \pi m_{\nu_s}} \frac{dN_{\rm{decay}}}{dE} D.
    \label{eq:flux}
\end{equation}
The flux $F$ per pixel is given by integrating $\Phi$ over the pixel solid angle $\Delta\Omega$.
The energy spectrum per decay is a delta function:
\begin{equation}
    \frac{dN_{\rm{decay}}}{dE} = \delta \left(E-\frac{m_{\nu_s}}{2} \right).
\label{eq:EnergySpectrum}
\end{equation}
The D-factor describes the dark matter density profile, $\rho$, of the Milky Way halo, integrated over the line of sight $s$ is given by
\begin{equation}
    D =  \int ds\, \rho(r(s,l,b)),
\end{equation}
where the radial distance from the Galactic center is described as 
\begin{equation}
    r=\sqrt{s^2+R^2-2sR\cos l \cos b},
\end{equation}
with $(l,b)$ the Galactic coordinates, and $R = 8.5~$kpc the distance from the Sun to the Galactic center. 
We consider both the spherically-symmetric Navarro-Frenk-White (NFW)~\cite{Navarro:1996gj} profile and a cored profile~\cite{10.1093/mnras/stw713}. The NFW profile is given by 
\begin{equation}
    \rho_{\rm NFW}(r) = \frac{\rho_s}{(r/r_s)(r/r_s+1)^2},
    \label{eq:NFW}
\end{equation}
parameterized by a scale radius $r_s$ and a characteristic density $\rho_s$.
The cored profile is given by 
\begin{equation}
    \rho_c(r) =f(r) \rho_{\rm NFW}(r)+\frac{1-f^2(r)}{4\pi r^2 r_c} M_{\rm NFW}(r),
\end{equation}
where the function, $f(r)=\tanh(r/r_c)$, considers the shallowness of the dark matter core below the core radius $r_c$. $M_{\rm NFW}(r)$ is the enclosed mass for the NFW density profile within $r$ that is given by
\begin{equation}
    M_{\rm NFW}(r) = M_{200} g_c \left[\ln\left(1+\frac{r}{r_s}\right)-\frac{r}{r_s}\left(1+\frac{r}{r_s}\right)^{-1} \right],
\end{equation}
where $M_{200} =1.11\times 10^{12} M_{\odot}$,  $g_c=1/\log(1+c)-c/(1+c)$, and $c=12.2$ is the halo concentration parameter~\cite{10.1093/mnras/stw713}. In our analysis we consider only complete core formation, reducing the number of parameters needed to specify the dark matter core properties. Adopting the parameters from \cite{PhysRevD.102.043012} for both NFW and cored profile, we set the core radius $r_c$ of the Milky Way to 1~kpc and use a scale radius of $r_s=26$~kpc. For the local dark matter density we take $\rho_0=0.28$~GeV~cm$^{-3}$, for which we find a density at the scale radius of $\rho_s=0.16$~GeV~cm$^{-3}$.

\subsubsection{Extragalactic sterile neutrino signal}
Additionally, decaying sterile neutrino contributes to the diffuse extragalactic signal, emitting at different redshifts. The average X-ray photon intensity is given as follows,
\begin{equation}
    \frac{d\Phi_{\rm{eg}}}{dE} = \frac{\Gamma_{\nu_s}\Omega_{\rm{DM}}\rho_c}{4 \pi m_{\nu_s}}\int_0^{\infty}\frac{dz}{H(z)}\delta\left(E(1+z)-\frac{m_{\nu_s}}{2}\right) , 
\end{equation}
with $H(z)=H_0\sqrt{\Omega_m(1+z)^3+\Omega_{\Lambda}}$ the Hubble parameter as a function of redshift $z$, and with $H_0=70 ~\rm km ~ s^{-1} ~ Mpc^{-1}$, $\Omega_m = 0.27$, $\Omega_{\Lambda} = 0.73$, $\Omega_{\rm{DM}}=0.22$, and the critical density $\rho_c=5.2\times10^{-6}\, \rm GeV ~cm^{-3}$. 

\begin{figure}[ht!]
\centering
\hskip7.mm
\includegraphics[width=0.48\textwidth]{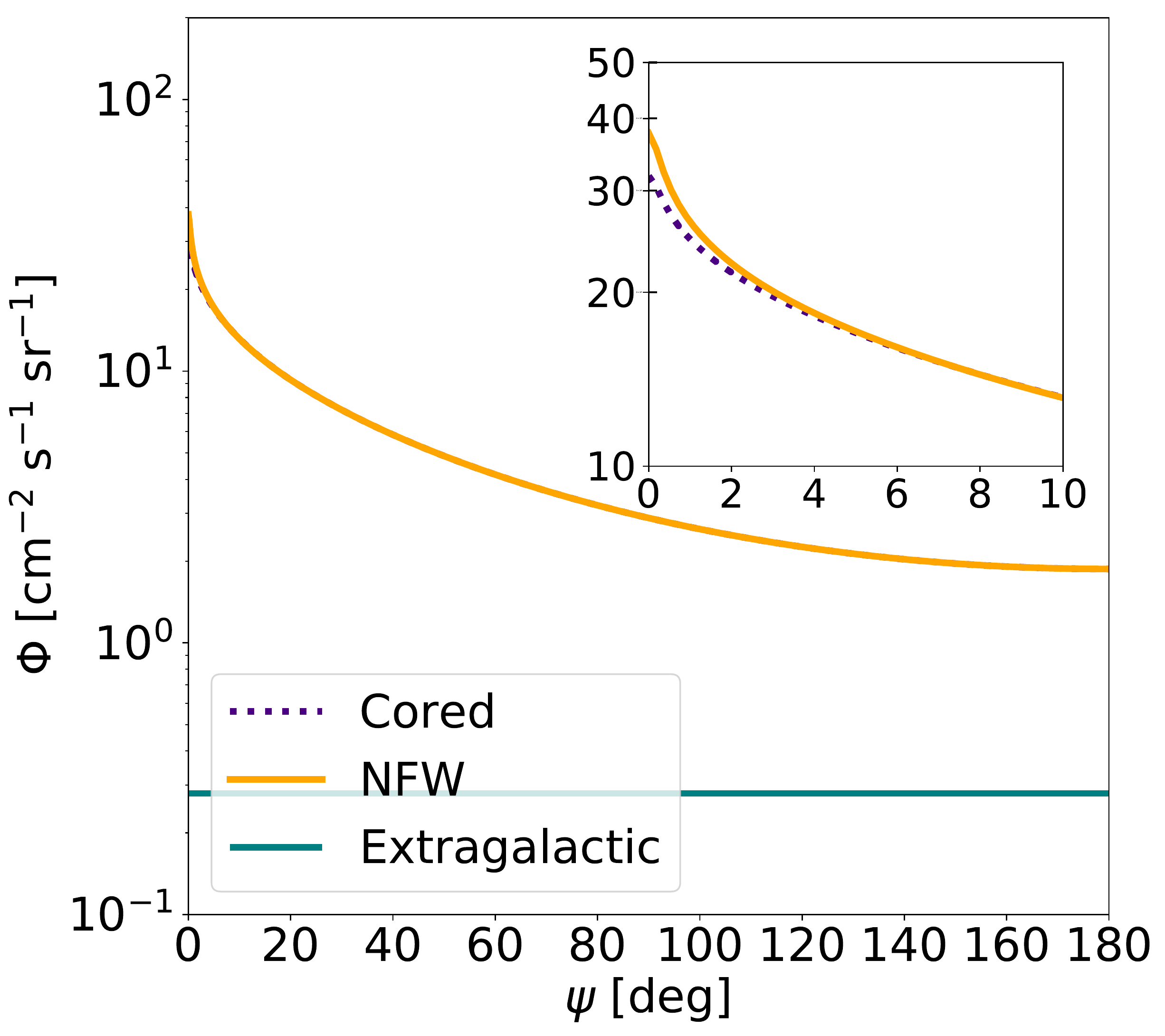}
\caption{The integrated flux as a function of the angle from the Galactic center $\psi$ for the Galactic component with both NFW (orange solid line) and cored profiles (purple dotted line), and the extragalactic component (green solid line). Sterile neutrino mass of 9~keV and energy range between 4.1 and 4.9~keV are adopted.}
\label{fig:profile_compare}
\end{figure}

Figure~\ref{fig:profile_compare} shows the integrated flux for the extragalactic (green solid line) and Galactic component by adopting NFW (orange solid line) and cored (purple dotted line) profiles as a function of angle $\Psi$ subtending from the Galactic center in the energy range of $[4.1:4.9]$~keV, with $m_{\nu_s}=9$~keV and $\Gamma_{\nu_s} = 10^{-28} ~{\rm s}^{-1}$. The Galactic flux associated with the cored profile is only slightly smaller with respect to that with the NFW profile at angles close to the Galactic center, and nearly identical further away. Furthermore, the extragalactic flux is more than an order of magnitude smaller than the Galactic flux within the small energy bins that we adopt in this paper and including the extragalactic flux will thus not lead to any significant improvement. 

\subsubsection{Axionlike particle signal}
ALPs can couple to several standard model particles. Here we consider the coupling to photons and electrons. In the case of a ALP-to-photon coupling, the decay into two photons produces a mono-energetic line at the energy of $m_a/2$, with a decay rate given by~\cite{Higaki:2014zua}
\begin{equation}
    \Gamma_{a\rightarrow \gamma\gamma} \simeq 5\times 10^{-29}  \left(\frac{m_a}{7~{\rm keV}}\right)^{3} \left(\frac{f_a}{5\times 10^{14}~{\rm GeV}}\right)^{-2} \, {\rm s}^{-1} ,
\end{equation}
where $m_a$ is the ALP mass and $f_a$ is the decay constant. One can convert the decay constant to the photon coupling $g_{a\gamma\gamma}$ through the following conversion
\begin{equation}
    f_{a}\equiv \frac{\alpha C_{a\gamma\gamma}}{2 \pi g_{a\gamma\gamma}},
\end{equation}
with $C_{a\gamma\gamma}=8/3 - 1.92\approx 0.75$~\cite{Irastorza:2018dyq}.

In order to explain the XENON1T excess by ALP, the ALP-photon coupling must be suppressed due to existing bounds~\cite{Irastorza:2018dyq}. We therefore consider ALPs that couple mainly to electrons, and assume an anomaly-free symmetry, where photons are only induced through threshold corrections. In this model, the decay rate is given by~\cite{Nakayama_2014}
\begin{equation}
    \Gamma_{a\rightarrow \gamma\gamma} \simeq 3.5\times 10^{-57} ~{\rm GeV} \left(\frac{m_a}{2~ {\rm keV}}\right)^7 \left(\frac{g_{ae}}{5\cdot 10^{-14}}\right)^{2},
\end{equation}
with $g_{ae}$ the coupling between the ALP and electron. 
The energy spectrum for the ALP is described by a delta function, as in equation~\ref{eq:EnergySpectrum}, where we replace $m_{\nu_s}$ by $m_a$. We further multiply the delta function by a factor of two to take into account that two photons are produced from the ALP decay. This allows for a direct comparison with the sterile neutrino flux and allows us to use the obtained X-ray bounds on the mixing angle and to convert to those on the coupling strength $g_{a\gamma\gamma}$ or $g_{ae}$.  It is therefore only necessary to construct sky maps for the sterile neutrino case, and the method for producing these maps is described in the following section.

\subsubsection{Sky maps}
The sky maps are generated with HEALPix\footnote{http://healpix.sf.net} using the software package \textit{healpy} \cite{Gorski:2004by,Zonca2019}, where we adopt its resolution parameter ${\rm Nside}=64$, which corresponds to a pixel size $\Delta \Omega = 0.84~{\rm deg}^2$. For each set of parameters $(m_{\nu_s},\Gamma_{\nu_s})$, and for each energy bin of width $\Delta E=E_2-E_1$, we obtain the expected number of X-ray photon counts from decaying sterile neutrinos coming from the galactic halo as well as extragalactic from a region on the sky $\Delta \Omega$ at position $(l,b)$ by 
\begin{equation}
    N(l,b) = T \int_{E_1}^{E_2} dE A_{\rm{eff}}(E) \int dE' P(E,E') \frac{dF}{dE'},
\end{equation}
where $T=2.5~$ks is the exposure time, $A_{\rm{eff}}(E)$ is the effective area and $P(E,E')$ takes into account the energy resolution of the detector. We apply the following normal distribution for the energy resolution,
\begin{equation}
    P(E,E') = \frac{1}{\sqrt{2\pi}\sigma_E}\exp\left[{-\frac{(E-E')^2}{2\sigma_E^2}}\right], 
\end{equation}
where $\sigma_E$ is related to the full width at half maximum through $\rm{FWHM}=2/\sqrt{2\ln2}\sigma_E$ with $\rm FWHM=138~$eV for eROSITA~\cite{merloni2012erosita}. We consider in total 13 energy bins around $m_{\nu_s}/2$ with bin size $\sigma_E$ for each sterile neutrino mass, and range the mass between $m_{\nu_s}[2:20]$~keV.

\subsubsection{Background events}
We consider an overall diffuse cosmic X-ray background (CXB), which is energy dependent and especially dominant at lower keV energies. 
It is represented by a power-law with photon index $\Gamma=1.42\pm 0.03$ and with a normalization at 1~keV of $8.44\pm 0.24$ photon cm$^{-2}$ s$^{-1}$ keV$^{-1}$ sr$^{-1}$~\cite{Lumb:2002sw}. 
Moreover, we consider eROSITA's detector background, which are high energy particles that show a flat spectral energy distribution with a normalization of $3.5\times 10^{-4}$ counts~keV$^{-1}$~s$^{-1}$~arcmin$^{-2}$~\cite{Predehl_2021}. We distribute both background contributions isotropically over the sky and apply the same energy binning as mentioned above. 

Besides the isotropic background contributions, X-ray bubbles are observed in the Milky Way, which are most prominent in the $0.6-1$~keV energy band and drop below the detector's background above $\sim 2.3$~keV~\cite{Predehl_2020}. An average count rate is measured by eROSITA in the $0.6-1$~keV energy band of 0.0038 photons s$^{-1}$ arcmin$^{-2}$ and 0.0026 photons s$^{-1}$ arcmin$^{-2}$ in the northern and southern bubbles respectively. 
We adopt a thermal spectrum to model the X-ray bubbles with a temperature of $0.3$~keV~\cite{Predehl_2020}, where we fix the normalization with the aforementioned count rates. For its morphology, we consider an uniform template of the Fermi bubbles, downloaded from 
\href{https://fermi.gsfc.nasa.gov/ssc/data/access/}{https://fermi.gsfc.nasa.gov/ssc/data/access/}. 

\begin{figure}[ht!]
    \centering
    \hskip7.mm
    \includegraphics[width=0.48\textwidth]{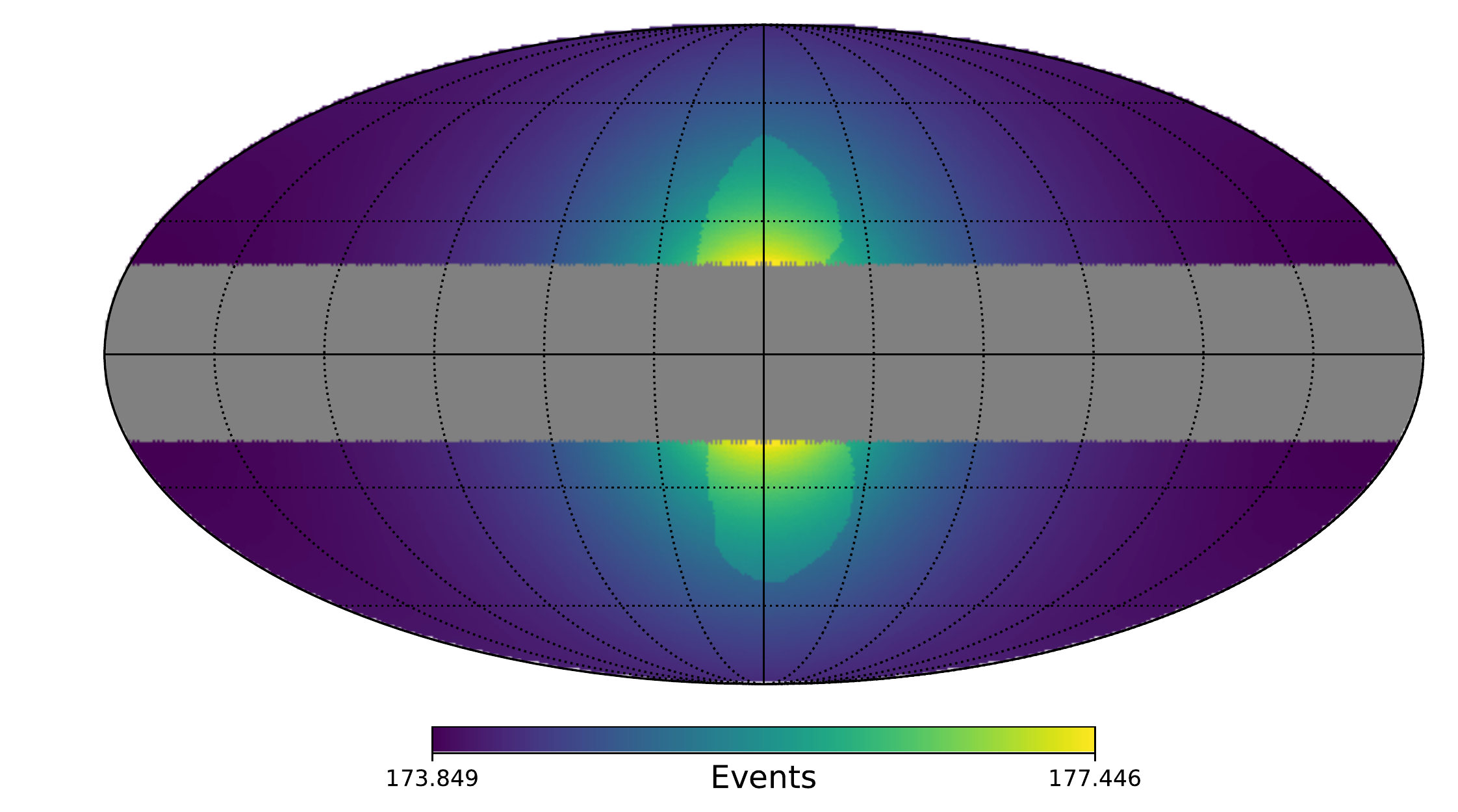}
\caption{X-ray sky map with sterile neutrino signal with $m_{\nu_s} = 9~{\rm keV}$, $\Gamma_{\nu_s} = 10^{-28} ~{\rm s}^{-1}$, as well as background components with 2.5~ks of eROSITA exposure within one energy bin around 4.5~keV whose width is $\sigma_E$. Pixels at the Galactic plane are removed with $\left| b \right|< 20^\circ$.}
    \label{fig:Skymap}
\end{figure}

Moreover, in order to exclude the extended emission from the Galactic plane, we remove all pixels with $\left| b \right|< 20^\circ$. 
Figure~\ref{fig:Skymap} illustrates a sky map with energy bin around $m_{\nu_s}=9~$keV with $\Gamma_{\nu_s}=10^{-28}\,\rm s^{-1}$, corresponding to $\sin^2(2\theta)\simeq 9.3\times 10^{-11}$, and additionally the isotropic and anisotropic background components. We analyze the full sky map under the signal hypothesis consisting of a sterile neutrino signal and background components, and under the null hypothesis with background components only.

\section{Analysis} \label{sec:anaysis}
We calculate the sensitivity to detect a sterile neutrino signal by performing a joint likelihood analysis on simulated data. We generate mock data sets assuming background only (the null hypothesis with $\Gamma = 0$), with Monte Carlo simulations following a Poisson distribution. This is performed for each pixel of our pixelized sky map in the binned energy window under consideration. 
For each sterile neutrino mass, we generate 500 mock data sets $n_i$, where $i$ runs over the energy bins as well as spatial pixels. The likelihood to obtain $n_i$ as a function of the decay rate for a specific sterile neutrino mass is given by the likelihood functions:
\begin{equation}
    \mathcal{L}(\Gamma) = \prod_i P\left[n_i|\mu_i(\Gamma)\right] = \prod_i \frac{\mu_i(\Gamma)^{n_i}e^{-\mu_i(\Gamma)}}{n_i!},
    \label{eq:likelihood}
\end{equation}
where $\mu_i(\Gamma)$ are the expected counts in each bin under the signal hypothesis with decaying dark matter and background component. 
The test statistic (TS) to determine the best-fit model under a maximum likelihood estimation is then defined as
\begin{equation}
    \text{TS} = -2\ln \left[ \frac{\mathcal{L}(\Gamma)}{\mathcal{L}_{\text{max}}}\right],
    \label{eq:TS}
\end{equation}
where $\mathcal{L}_{\text{max}}$ is the the maximum likelihood. We obtain upper limits on the decay rates at 95\% confidence level (CL), which corresponds to a test statistic of $\rm TS = 2.71$.

\section{Results and discussion}\label{sec:results}
\begin{figure}[ht!]
    \centering
    \hskip7.mm
    \includegraphics[width=0.48\textwidth]{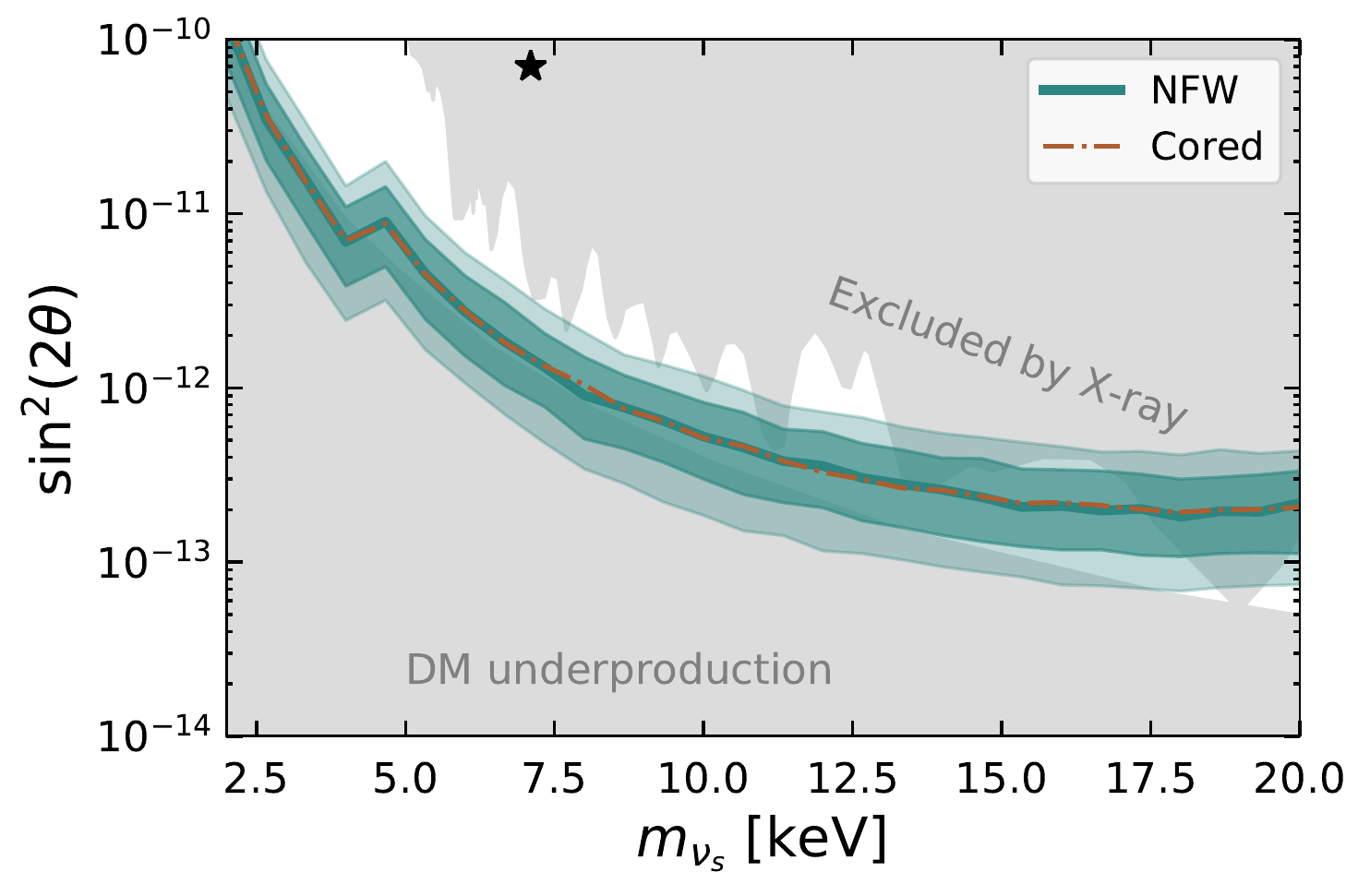}
\caption{Sensitivity to the mixing angle as a function of the sterile neutrino mass for the cored (dashed orange) and NFW (green) profiles. The green bands show the 68\% and 95\% containment regions of the sensitivities and the median (solid line) from the Monte Carlo runs with an NFW profile. 
The black star indicates the best-fit for the unidentified $3.5$~keV line with mixing angle $\sin^2(2\theta)\simeq (0.2-2)\times 10^{-10}$~\cite{Bulbul_2014}, the upper grey area current X-ray constraints~\cite{Horiuchi:2013noa, Ng:2015gfa, Perez:2016tcq, Ng:2019gch,Abazajian_2017,Caputo:2019djj,Roach_2020,foster2021deep}, while the lower grey area indicates the theoretical upper bound for dark matter underproduction~\cite{Serpico_2005,Cherry_2017}.}
\label{fig:Sensitivity_mixinganlge}
\end{figure}

We analyse the simulated sky maps under the null hypothesis, and report the sensitivity of eROSITA on the mixing angle as a function of the sterile neutrino mass. Figure~\ref{fig:Sensitivity_mixinganlge} shows the result by applying the NFW profiles by removing the Galactic plane with $|b|<20^\circ$. The two bands show the 68\% and 95\% containment regions from the Monte Carlo runs, while the solid line represents the median. The upper grey area represents limits based on current X-ray observations~\cite{Horiuchi:2013noa, Ng:2015gfa, Perez:2016tcq, Ng:2019gch,Abazajian_2017,Caputo:2019djj,Roach_2020,foster2021deep}, while the lower grey area represents the theoretical lower limit for dark matter underproduction~\cite{Serpico_2005,Cherry_2017}. With eROSITA, we will nearly close the gap between current lower and upper bounds. Reference~\cite{Barinov_2021} obtains similar estimates at lower sterile neutrino masses by analyzing the signal from the inner $60^\circ$ region around the Galactic center with eROSITA, whereas our estimates are stronger at larger sterile neutrino masses due to the larger region of interest as well as the removal of the Galactic plane. 
Moreover, we indicate the best-fit of the unidentified $3.5$~keV line by Ref.~\cite{Bulbul_2014} as a black star, with mass $m_{\nu_s}=7.1$~keV and mixing angle $\sin^2(2\theta)=7\times 10^{-11}$. With an exposure time of $T=2.5$~ks, eROSITA will be sensitive to the $3.5$~keV line and can even constrain the mixing angles up to nearly two orders of magnitude lower than the best-fit at $m_{\nu_s}=7.1$~keV. 

\begin{figure}[ht!]
    \centering
    \hskip7.mm
    \includegraphics[width=0.48\textwidth]{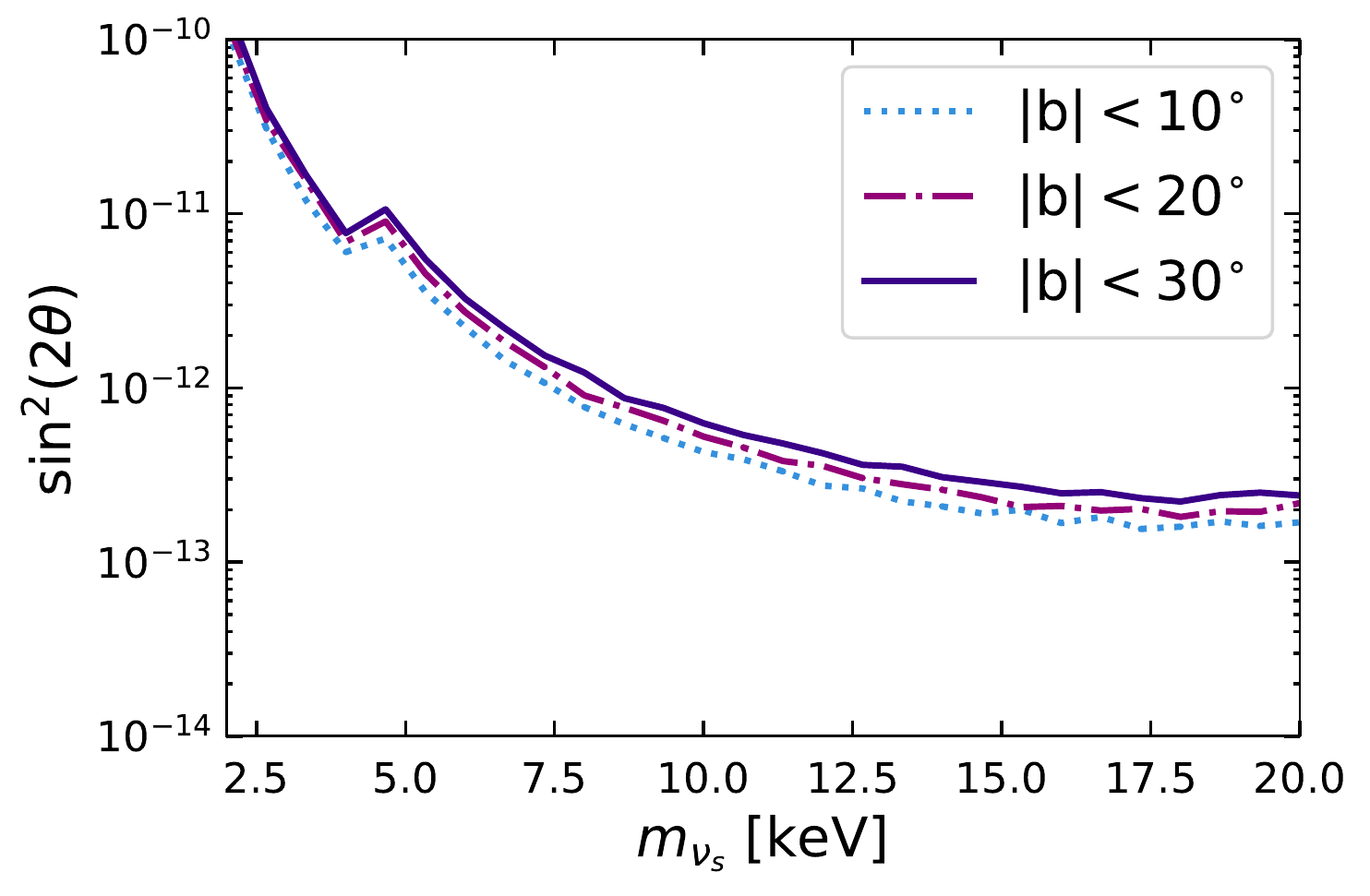}
\caption{Sensitivity to the mixing angle, adopting a NFW density profile. The pixels with the following Galactic latitudes are removed: $|b|<10^\circ$ (dotted), $|b|<20^\circ$ (dashed dotted) and $|b|<30^\circ$ (solid). }
\label{fig:Sensitivity_mixinganlge2}
\end{figure}

We evaluate how sensitive these results are to some aspects in our analysis.
The median of the Monte Carlo runs with a cored profile is illustrated as the orange dashed dotted line, and by comparing with the NFW profile, we find little dependence on the density profiles. Indeed, figure~\ref{fig:profile_compare} shows that the difference between the density profiles is most prominent at the inner regions, which we exclude with the $|b|<20^\circ$ cut. Additionally, we test the impact of excluding the Galactic plane at different latitudes, as shown in Fig.~\ref{fig:Sensitivity_mixinganlge2}, after removing the galactic latitudes $|b|<10^\circ$ (dashed), $|b|<20^\circ$ (dashed dotted) and $|b|<30^\circ$ (solid line) for a NFW density profile. As expected, including a larger latitude $|b|<10^\circ$, shows slightly stronger constraints with respect to $|b|<30^\circ$, however only by a factor of $\sim1.2$. The X-ray bubbles contribute to the lowest energies we consider. However, since the sky coverage of the Fermi bubbles template is only $7\%$ (before masking), we find that our limits weaken only by a factor 1.1 at most by including the X-ray bubbles.

\begin{figure}[ht!]
    \centering
    \hskip7.mm
    \includegraphics[width=0.48\textwidth]{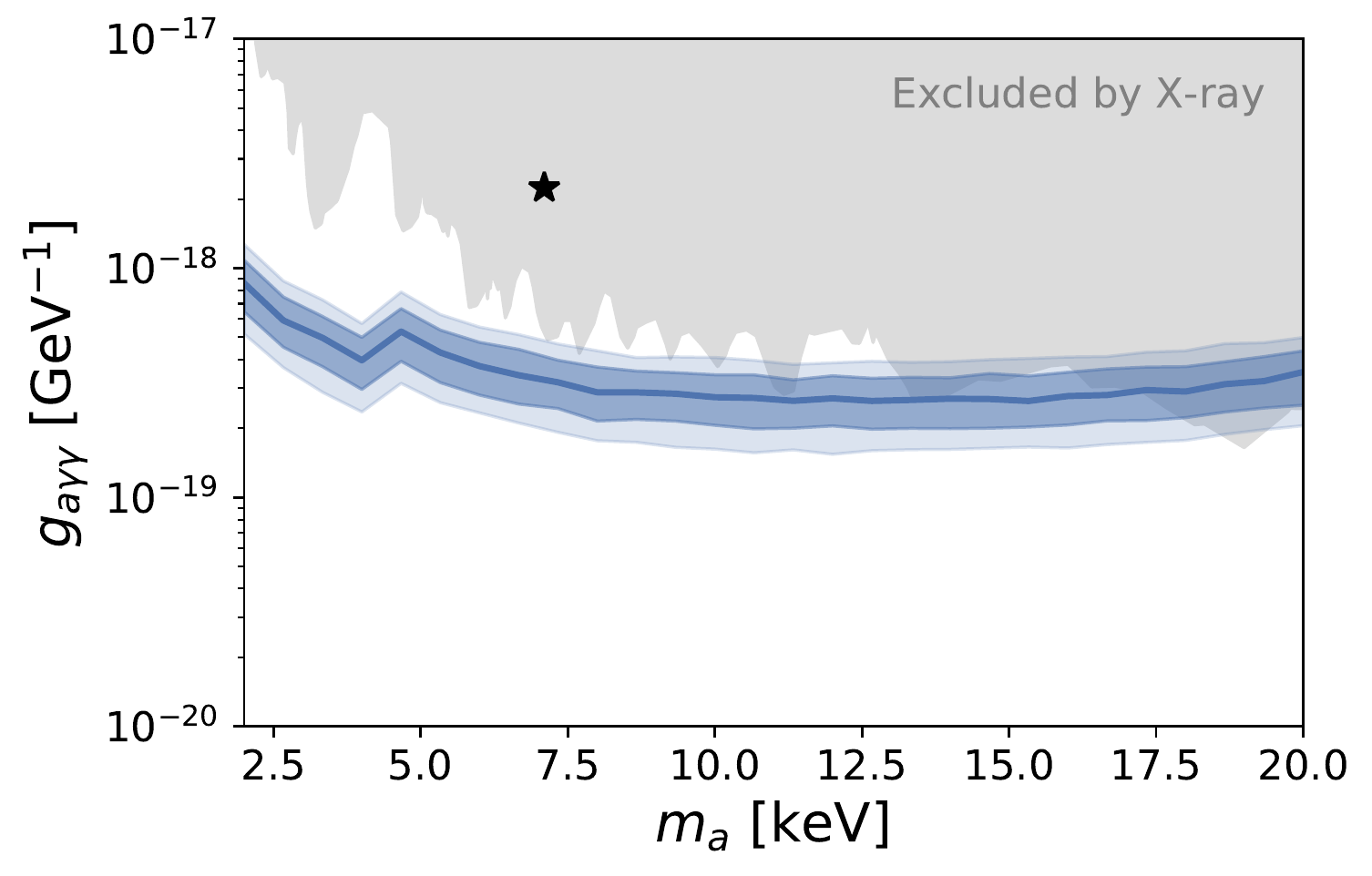}
\caption{Current and future limits on the ALP photon coupling. The blue bands show the 68\% and 95\% containment regions of sensitivities from the Monte Carlo runs and the blue solid line the median. The grey shaded area shows the region that is excluded by current X-ray observations~\cite{Horiuchi:2013noa, Ng:2015gfa, Perez:2016tcq, Ng:2019gch,Abazajian_2017,Caputo:2019djj,Roach_2020,foster2021deep}, and the black star indicates the best-fit from the unidentified 3.5~keV line with $g_{a\gamma \gamma}\simeq(0.6-2)\times10^{-18}$~\cite{Bulbul_2014}.} 
\label{fig:ALP_coupling_photon} 
\end{figure}

The bounds on the mixing angle can be converted to the ALP-photon coupling $g_{a\gamma\gamma}$ and ALP-electron coupling $g_{ae}$, 
and the sensitivity of eROSITA to the photon coupling is shown in Fig.~\ref{fig:ALP_coupling_photon}. Again, the best-fit for the 3.5~keV line is indicated by a black star, and with future eROSITA observations, an ALP scenario can be probed as an explanation for the unidentified X-ray line. Even though recent work found no evidence for an unassociated X-ray line~\cite{foster2021deep}, in which the current X-ray limits are shown as the grey shaded area (see ~\cite{Horiuchi:2013noa, Ng:2015gfa, Perez:2016tcq, Ng:2019gch,Abazajian_2017,Caputo:2019djj,Roach_2020,foster2021deep}), eROSITA will be able to probe a region of the parameter space not yet excluded by current X-ray limits.

\begin{figure}[ht!]
    \centering
    \hskip7.mm
    \includegraphics[width=0.48\textwidth]{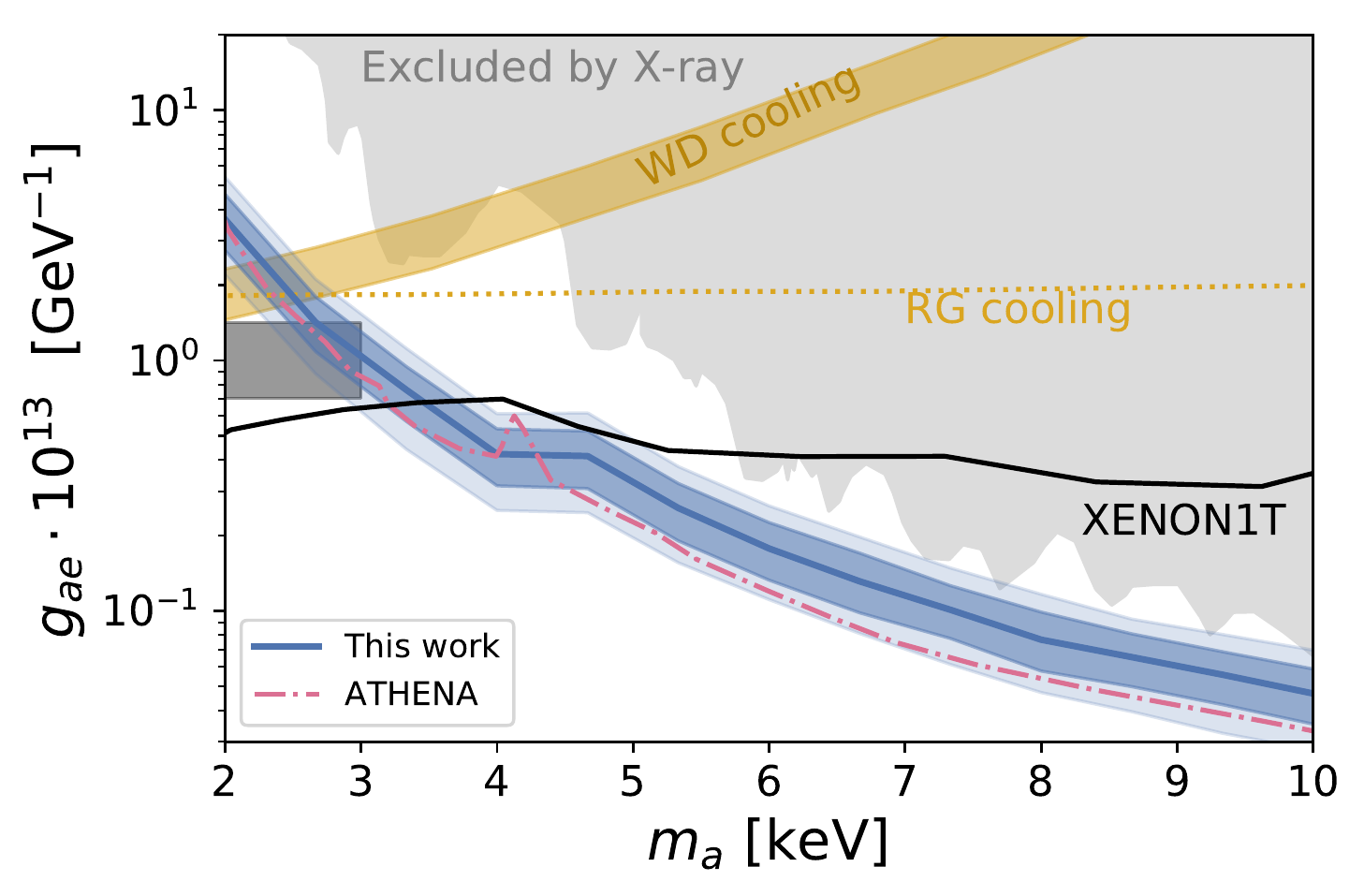}
\caption{Current and future limits on the ALP electron coupling. The blue bands show the 1 and $2\sigma$ sensitivities from the Monte Carlo runs of this work and the blue solid line the median. Moreover, the black solid line represents the XENON1T limit at 90\% C.L.~\cite{Aprile_2020} and the dark grey box highlights roughly the best-fit parameter space from Ref.~\cite{Takahashi:2020bpq}. The red dashed line is the expected sensitivity of Athena~\cite{Neronov_2016_2}. The grey shaded area shows the region that is excluded by current X-ray observations~\cite{Horiuchi:2013noa, Ng:2015gfa, Perez:2016tcq, Ng:2019gch,Abazajian_2017,Caputo:2019djj,Roach_2020,foster2021deep}. The yellow dotted line represents the preferred values to explain the red giant cooling anomaly~\cite{Takahashi:2020bpq,Viaux_2013} and the yellow shaded area the white dwarf cooling anomaly~\cite{Takahashi:2020bpq,Bertolami:2014wua}.}
\label{fig:ALP_coupling_electron}
\end{figure}

Furthermore, we consider an anomaly-free ALP model in order to explain the XENON1T excess, an excess that has been observed to be most prominent at ALP mass of $m_a=2$--3~keV and electron coupling $g_{ae}\sim10^{-13}$, and has the best fit at $m_a=2.3$~keV~\cite{Aprile_2020}. We test if eROSITA will be able to confirm this, and we show its expected sensitivity in Fig.~\ref{fig:ALP_coupling_electron}, where the two bands show as before the 68\% and 95\% containment bands from the Monte Carlo runs, while the solid line represents the median. 
We show the region that is excluded by current X-ray observations in grey, and with eROSITA we can indeed probe a parameter space not yet constrained~\cite{Horiuchi:2013noa, Ng:2015gfa, Perez:2016tcq, Ng:2019gch,Abazajian_2017,Caputo:2019djj,Roach_2020,foster2021deep}. 
The black solid line represents the XENON1T limits (note that they are however given at 90\% CL)~\cite{Aprile_2020}. The XENON1T excess best-fit may not be reached by future eROSITA data, however, if the best-fit alters towards $m_a\sim 3.5$~keV, which is still inside the XENON1T excess region of interest with energies between 1--7~keV, an ALP origin could be confirmed. Interestingly, the expected sensitivity of Athena taken from Ref.~\cite{Neronov_2016_2} show comparable sensitivity. 

Moreover, the stellar cooling anomaly can be explained by an ALP contribution. The preferred region for the white dwarf cooling anomaly is illustrated as the yellow shaded area, while the preferred values for the red giant branch in globular clusters is illustrated as the yellow dotted line~\cite{Takahashi:2020bpq,Bertolami:2014wua,Viaux_2013}, and the preferred regions are close to the XENON1T excess. Reference~\cite{Takahashi:2020bpq} points out that the stellar cooling argument and the XENON1T excess cross each other for ALP constituting only a fraction $r\simeq0.1$ of the total dark matter, since the XENONT1T data scales as $1/\sqrt{r}$. The excess could possibly be explained by a combination of ALP and another background component like tritium, as suggested by~\cite{Aprile_2020}.

\section{conclusion} \label{sec:conclusion}
In this work, we search for a decaying sterile neutrino and axionlike-particle (ALP) signal from the Galactic halo with all-sky eROSITA survey after four years observations. We generate mock data sets with the diffuse cosmic X-ray background and eROSITA's detector background, as well as expected sky maps with counts from decaying sterile neutrinos in the Galactic halo. By performing a likelihood analysis, we set stringent bounds on the mixing angle of sterile neutrinos. We also convert these bounds to the ALP-photon coupling ($g_{a\gamma\gamma}$) and ALP-electron coupling ($g_{ae}$). We consider a cored and cusped Navarro-Frenk-White density profiles and find only tiny dependence on the choice of the density profile. 

We will be able to probe a value for the mixing angle of sterile neutrinos up to nearly two orders of magnitude below the best-fit value that could explain the unidentified 3.5~keV line~\cite{Bulbul_2014} and one order of magnitude stronger than the existing upper limits claimed in the literature~\cite{Horiuchi:2013noa, Ng:2015gfa, Perez:2016tcq, Ng:2019gch,Abazajian_2017,Caputo:2019djj,Roach_2020,foster2021deep}. 
In an accompanying paper~\cite{ando2021decaying}, a similar analysis is performed discussing the eROSITA sensitivity to sterile neutrino decay based on analyzing Milky-Way satellite galaxies.

We will also be able to probe a large parameter space for the ALP couplings to photons and electrons that are not yet excluded by X-ray observations, to the same degree of improvement as in the case of sterile neutrinos. We investigate both a generic model for the ALP-to-photon coupling and a more specific anomaly-free symmetry that has been proposed to explain the XENON1T excess of electron recoil events~\cite{Aprile_2020}. Indeed, the XENON1T excess could possibly be explained by an ALP origin for an excess at $m_a\sim 3$~keV, which might be well tested with eROSITA.

We note that in estimating the sensitivity, all detector and astrophysical lines are neglected. It is understood that the sensitivity at these line energies would decrease significantly due to signal-background degeneracy. Near-future high-energy-resolution detectors, such as Athena and XRISM, may alleviate this by performing line diagnostics analysis based on different line shifts between signal and backgrounds~\cite{Speckhard:2015eva, Powell:2016zbo, Zhong:2020wre}.

\begin{acknowledgments}
We thank Marco Chianese and Oscar Macias for helpful discussions. This works was supported by Institute of Physics at the University of Amsterdam, and SA acknowledges the support by JSPS/MEXT KAKENHI Grant Numbers JP17H04836, JP18H04340, JP20H05850, and JP20H05861 (SA).

\end{acknowledgments}

\bibliography{references.bib}

\begin{thebibliography}{71}%
\makeatletter
\providecommand \@ifxundefined [1]{%
 \@ifx{#1\undefined}
}%
\providecommand \@ifnum [1]{%
 \ifnum #1\expandafter \@firstoftwo
 \else \expandafter \@secondoftwo
 \fi
}%
\providecommand \@ifx [1]{%
 \ifx #1\expandafter \@firstoftwo
 \else \expandafter \@secondoftwo
 \fi
}%
\providecommand \natexlab [1]{#1}%
\providecommand \enquote  [1]{``#1''}%
\providecommand \bibnamefont  [1]{#1}%
\providecommand \bibfnamefont [1]{#1}%
\providecommand \citenamefont [1]{#1}%
\providecommand \href@noop [0]{\@secondoftwo}%
\providecommand \href [0]{\begingroup \@sanitize@url \@href}%
\providecommand \@href[1]{\@@startlink{#1}\@@href}%
\providecommand \@@href[1]{\endgroup#1\@@endlink}%
\providecommand \@sanitize@url [0]{\catcode `\\12\catcode `\$12\catcode
  `\&12\catcode `\#12\catcode `\^12\catcode `\_12\catcode `\%12\relax}%
\providecommand \@@startlink[1]{}%
\providecommand \@@endlink[0]{}%
\providecommand \url  [0]{\begingroup\@sanitize@url \@url }%
\providecommand \@url [1]{\endgroup\@href {#1}{\urlprefix }}%
\providecommand \urlprefix  [0]{URL }%
\providecommand \Eprint [0]{\href }%
\providecommand \doibase [0]{http://dx.doi.org/}%
\providecommand \selectlanguage [0]{\@gobble}%
\providecommand \bibinfo  [0]{\@secondoftwo}%
\providecommand \bibfield  [0]{\@secondoftwo}%
\providecommand \translation [1]{[#1]}%
\providecommand \BibitemOpen [0]{}%
\providecommand \bibitemStop [0]{}%
\providecommand \bibitemNoStop [0]{.\EOS\space}%
\providecommand \EOS [0]{\spacefactor3000\relax}%
\providecommand \BibitemShut  [1]{\csname bibitem#1\endcsname}%
\let\auto@bib@innerbib\@empty
\bibitem [{\citenamefont {Frenk}\ and\ \citenamefont
  {White}(2012)}]{Frenk_2012}%
  \BibitemOpen
  \bibfield  {author} {\bibinfo {author} {\bibfnamefont {C.}~\bibnamefont
  {Frenk}}\ and\ \bibinfo {author} {\bibfnamefont {S.}~\bibnamefont {White}},\
  }\href {\doibase 10.1002/andp.201200212} {\bibfield  {journal} {\bibinfo
  {journal} {Annalen der Physik}\ }\textbf {\bibinfo {volume} {524}},\ \bibinfo
  {pages} {507–534} (\bibinfo {year} {2012})}\BibitemShut {NoStop}%
\bibitem [{\citenamefont {Roszkowski}\ \emph {et~al.}(2018)\citenamefont
  {Roszkowski}, \citenamefont {Sessolo},\ and\ \citenamefont
  {Trojanowski}}]{Roszkowski_2018}%
  \BibitemOpen
  \bibfield  {author} {\bibinfo {author} {\bibfnamefont {L.}~\bibnamefont
  {Roszkowski}}, \bibinfo {author} {\bibfnamefont {E.~M.}\ \bibnamefont
  {Sessolo}}, \ and\ \bibinfo {author} {\bibfnamefont {S.}~\bibnamefont
  {Trojanowski}},\ }\href {\doibase 10.1088/1361-6633/aab913} {\bibfield
  {journal} {\bibinfo  {journal} {Reports on Progress in Physics}\ }\textbf
  {\bibinfo {volume} {81}},\ \bibinfo {pages} {066201} (\bibinfo {year}
  {2018})}\BibitemShut {NoStop}%
\bibitem [{\citenamefont {Silk}\ \emph {et~al.}(2010)\citenamefont {Silk} \emph
  {et~al.}}]{Bertone:2010zza}%
  \BibitemOpen
  \bibfield  {author} {\bibinfo {author} {\bibfnamefont {J.}~\bibnamefont
  {Silk}} \emph {et~al.},\ }\href {\doibase 10.1017/CBO9780511770739} {\emph
  {\bibinfo {title} {{Particle Dark Matter: Observations, Models and
  Searches}}}},\ edited by\ \bibinfo {editor} {\bibfnamefont {G.}~\bibnamefont
  {Bertone}}\ (\bibinfo  {publisher} {Cambridge Univ. Press},\ \bibinfo
  {address} {Cambridge},\ \bibinfo {year} {2010})\BibitemShut {NoStop}%
\bibitem [{\citenamefont {Bode}\ \emph {et~al.}(2001)\citenamefont {Bode},
  \citenamefont {Ostriker},\ and\ \citenamefont {Turok}}]{Bode_2001}%
  \BibitemOpen
  \bibfield  {author} {\bibinfo {author} {\bibfnamefont {P.}~\bibnamefont
  {Bode}}, \bibinfo {author} {\bibfnamefont {J.~P.}\ \bibnamefont {Ostriker}},
  \ and\ \bibinfo {author} {\bibfnamefont {N.}~\bibnamefont {Turok}},\ }\href
  {\doibase 10.1086/321541} {\bibfield  {journal} {\bibinfo  {journal} {The
  Astrophysical Journal}\ }\textbf {\bibinfo {volume} {556}},\ \bibinfo {pages}
  {93} (\bibinfo {year} {2001})}\BibitemShut {NoStop}%
\bibitem [{\citenamefont {Shi}\ and\ \citenamefont
  {Fuller}(1999)}]{PhysRevLett.82.2832}%
  \BibitemOpen
  \bibfield  {author} {\bibinfo {author} {\bibfnamefont {X.}~\bibnamefont
  {Shi}}\ and\ \bibinfo {author} {\bibfnamefont {G.~M.}\ \bibnamefont
  {Fuller}},\ }\href {\doibase 10.1103/PhysRevLett.82.2832} {\bibfield
  {journal} {\bibinfo  {journal} {Phys. Rev. Lett.}\ }\textbf {\bibinfo
  {volume} {82}},\ \bibinfo {pages} {2832} (\bibinfo {year}
  {1999})}\BibitemShut {NoStop}%
\bibitem [{\citenamefont {Abazajian}\ \emph {et~al.}(2001)\citenamefont
  {Abazajian}, \citenamefont {Fuller},\ and\ \citenamefont
  {Patel}}]{PhysRevD.64.023501}%
  \BibitemOpen
  \bibfield  {author} {\bibinfo {author} {\bibfnamefont {K.}~\bibnamefont
  {Abazajian}}, \bibinfo {author} {\bibfnamefont {G.~M.}\ \bibnamefont
  {Fuller}}, \ and\ \bibinfo {author} {\bibfnamefont {M.}~\bibnamefont
  {Patel}},\ }\href {\doibase 10.1103/PhysRevD.64.023501} {\bibfield  {journal}
  {\bibinfo  {journal} {Phys. Rev. D}\ }\textbf {\bibinfo {volume} {64}},\
  \bibinfo {pages} {023501} (\bibinfo {year} {2001})}\BibitemShut {NoStop}%
\bibitem [{\citenamefont {Dolgov}\ and\ \citenamefont
  {Hansen}(2002)}]{Dolgov:2000ew}%
  \BibitemOpen
  \bibfield  {author} {\bibinfo {author} {\bibfnamefont {A.~D.}\ \bibnamefont
  {Dolgov}}\ and\ \bibinfo {author} {\bibfnamefont {S.~H.}\ \bibnamefont
  {Hansen}},\ }\href {\doibase 10.1016/S0927-6505(01)00115-3} {\bibfield
  {journal} {\bibinfo  {journal} {Astropart. Phys.}\ }\textbf {\bibinfo
  {volume} {16}},\ \bibinfo {pages} {339} (\bibinfo {year} {2002})},\ \Eprint
  {http://arxiv.org/abs/hep-ph/0009083} {arXiv:hep-ph/0009083} \BibitemShut
  {NoStop}%
\bibitem [{\citenamefont {Canetti}\ \emph {et~al.}(2013)\citenamefont
  {Canetti}, \citenamefont {Drewes}, \citenamefont {Frossard},\ and\
  \citenamefont {Shaposhnikov}}]{Canetti:2012kh}%
  \BibitemOpen
  \bibfield  {author} {\bibinfo {author} {\bibfnamefont {L.}~\bibnamefont
  {Canetti}}, \bibinfo {author} {\bibfnamefont {M.}~\bibnamefont {Drewes}},
  \bibinfo {author} {\bibfnamefont {T.}~\bibnamefont {Frossard}}, \ and\
  \bibinfo {author} {\bibfnamefont {M.}~\bibnamefont {Shaposhnikov}},\ }\href
  {\doibase 10.1103/PhysRevD.87.093006} {\bibfield  {journal} {\bibinfo
  {journal} {Phys. Rev. D}\ }\textbf {\bibinfo {volume} {87}},\ \bibinfo
  {pages} {093006} (\bibinfo {year} {2013})},\ \Eprint
  {http://arxiv.org/abs/1208.4607} {arXiv:1208.4607 [hep-ph]} \BibitemShut
  {NoStop}%
\bibitem [{\citenamefont {Boyarsky}\ \emph {et~al.}(2019)\citenamefont
  {Boyarsky}, \citenamefont {Drewes}, \citenamefont {Lasserre}, \citenamefont
  {Mertens},\ and\ \citenamefont {Ruchayskiy}}]{Boyarsky:2018tvu}%
  \BibitemOpen
  \bibfield  {author} {\bibinfo {author} {\bibfnamefont {A.}~\bibnamefont
  {Boyarsky}}, \bibinfo {author} {\bibfnamefont {M.}~\bibnamefont {Drewes}},
  \bibinfo {author} {\bibfnamefont {T.}~\bibnamefont {Lasserre}}, \bibinfo
  {author} {\bibfnamefont {S.}~\bibnamefont {Mertens}}, \ and\ \bibinfo
  {author} {\bibfnamefont {O.}~\bibnamefont {Ruchayskiy}},\ }\href {\doibase
  10.1016/j.ppnp.2018.07.004} {\bibfield  {journal} {\bibinfo  {journal} {Prog.
  Part. Nucl. Phys.}\ }\textbf {\bibinfo {volume} {104}},\ \bibinfo {pages} {1}
  (\bibinfo {year} {2019})},\ \Eprint {http://arxiv.org/abs/1807.07938}
  {arXiv:1807.07938 [hep-ph]} \BibitemShut {NoStop}%
\bibitem [{\citenamefont {Kajita}(1999)}]{Kajita_1999}%
  \BibitemOpen
  \bibfield  {author} {\bibinfo {author} {\bibfnamefont {T.}~\bibnamefont
  {Kajita}},\ }\href {\doibase 10.1016/s0920-5632(99)00407-7} {\bibfield
  {journal} {\bibinfo  {journal} {Nuclear Physics B - Proceedings Supplements}\
  }\textbf {\bibinfo {volume} {77}},\ \bibinfo {pages} {123–132} (\bibinfo
  {year} {1999})}\BibitemShut {NoStop}%
\bibitem [{\citenamefont {Asaka}\ \emph {et~al.}(2005)\citenamefont {Asaka},
  \citenamefont {Blanchet},\ and\ \citenamefont {Shaposhnikov}}]{Asaka_2005}%
  \BibitemOpen
  \bibfield  {author} {\bibinfo {author} {\bibfnamefont {T.}~\bibnamefont
  {Asaka}}, \bibinfo {author} {\bibfnamefont {S.}~\bibnamefont {Blanchet}}, \
  and\ \bibinfo {author} {\bibfnamefont {M.}~\bibnamefont {Shaposhnikov}},\
  }\href {\doibase 10.1016/j.physletb.2005.09.070} {\bibfield  {journal}
  {\bibinfo  {journal} {Physics Letters B}\ }\textbf {\bibinfo {volume}
  {631}},\ \bibinfo {pages} {151–156} (\bibinfo {year} {2005})}\BibitemShut
  {NoStop}%
\bibitem [{\citenamefont {Boyarsky}\ \emph {et~al.}(2009)\citenamefont
  {Boyarsky}, \citenamefont {Ruchayskiy},\ and\ \citenamefont
  {Shaposhnikov}}]{Boyarsky_2009}%
  \BibitemOpen
  \bibfield  {author} {\bibinfo {author} {\bibfnamefont {A.}~\bibnamefont
  {Boyarsky}}, \bibinfo {author} {\bibfnamefont {O.}~\bibnamefont
  {Ruchayskiy}}, \ and\ \bibinfo {author} {\bibfnamefont {M.}~\bibnamefont
  {Shaposhnikov}},\ }\href {\doibase 10.1146/annurev.nucl.010909.083654}
  {\bibfield  {journal} {\bibinfo  {journal} {Annual Review of Nuclear and
  Particle Science}\ }\textbf {\bibinfo {volume} {59}},\ \bibinfo {pages}
  {191–214} (\bibinfo {year} {2009})}\BibitemShut {NoStop}%
\bibitem [{\citenamefont {Sicilian}\ \emph {et~al.}(2020)\citenamefont
  {Sicilian}, \citenamefont {Cappelluti}, \citenamefont {Bulbul}, \citenamefont
  {Civano}, \citenamefont {Moscetti},\ and\ \citenamefont
  {Reynolds}}]{Sicilian:2020glg}%
  \BibitemOpen
  \bibfield  {author} {\bibinfo {author} {\bibfnamefont {D.}~\bibnamefont
  {Sicilian}}, \bibinfo {author} {\bibfnamefont {N.}~\bibnamefont
  {Cappelluti}}, \bibinfo {author} {\bibfnamefont {E.}~\bibnamefont {Bulbul}},
  \bibinfo {author} {\bibfnamefont {F.}~\bibnamefont {Civano}}, \bibinfo
  {author} {\bibfnamefont {M.}~\bibnamefont {Moscetti}}, \ and\ \bibinfo
  {author} {\bibfnamefont {C.~S.}\ \bibnamefont {Reynolds}},\ }\href@noop {} {\
   (\bibinfo {year} {2020})},\ \Eprint {http://arxiv.org/abs/2008.02283}
  {arXiv:2008.02283 [astro-ph.HE]} \BibitemShut {NoStop}%
\bibitem [{\citenamefont {Abazajian}(2021)}]{abazajian2021neutrinos}%
  \BibitemOpen
  \bibfield  {author} {\bibinfo {author} {\bibfnamefont {K.~N.}\ \bibnamefont
  {Abazajian}},\ }\href@noop {} {\enquote {\bibinfo {title} {Neutrinos in
  astrophysics and cosmology: Theoretical advanced study institute (tasi) 2020
  lectures},}\ } (\bibinfo {year} {2021}),\ \Eprint
  {http://arxiv.org/abs/2102.10183} {arXiv:2102.10183 [hep-ph]} \BibitemShut
  {NoStop}%
\bibitem [{\citenamefont {Bulbul}\ \emph {et~al.}(2014)\citenamefont {Bulbul},
  \citenamefont {Markevitch}, \citenamefont {Foster}, \citenamefont {Smith},
  \citenamefont {Loewenstein},\ and\ \citenamefont {Randall}}]{Bulbul_2014}%
  \BibitemOpen
  \bibfield  {author} {\bibinfo {author} {\bibfnamefont {E.}~\bibnamefont
  {Bulbul}}, \bibinfo {author} {\bibfnamefont {M.}~\bibnamefont {Markevitch}},
  \bibinfo {author} {\bibfnamefont {A.}~\bibnamefont {Foster}}, \bibinfo
  {author} {\bibfnamefont {R.~K.}\ \bibnamefont {Smith}}, \bibinfo {author}
  {\bibfnamefont {M.}~\bibnamefont {Loewenstein}}, \ and\ \bibinfo {author}
  {\bibfnamefont {S.~W.}\ \bibnamefont {Randall}},\ }\href {\doibase
  10.1088/0004-637x/789/1/13} {\bibfield  {journal} {\bibinfo  {journal} {The
  Astrophysical Journal}\ }\textbf {\bibinfo {volume} {789}},\ \bibinfo {pages}
  {13} (\bibinfo {year} {2014})}\BibitemShut {NoStop}%
\bibitem [{\citenamefont {Boyarsky}\ \emph {et~al.}(2014)\citenamefont
  {Boyarsky}, \citenamefont {Ruchayskiy}, \citenamefont {Iakubovskyi},\ and\
  \citenamefont {Franse}}]{Boyarsky_2014}%
  \BibitemOpen
  \bibfield  {author} {\bibinfo {author} {\bibfnamefont {A.}~\bibnamefont
  {Boyarsky}}, \bibinfo {author} {\bibfnamefont {O.}~\bibnamefont
  {Ruchayskiy}}, \bibinfo {author} {\bibfnamefont {D.}~\bibnamefont
  {Iakubovskyi}}, \ and\ \bibinfo {author} {\bibfnamefont {J.}~\bibnamefont
  {Franse}},\ }\href {\doibase 10.1103/physrevlett.113.251301} {\bibfield
  {journal} {\bibinfo  {journal} {Physical Review Letters}\ }\textbf {\bibinfo
  {volume} {113}} (\bibinfo {year} {2014}),\
  10.1103/physrevlett.113.251301}\BibitemShut {NoStop}%
\bibitem [{\citenamefont {Boyarsky}\ \emph {et~al.}(2015)\citenamefont
  {Boyarsky}, \citenamefont {Franse}, \citenamefont {Iakubovskyi},\ and\
  \citenamefont {Ruchayskiy}}]{PhysRevLett.115.161301}%
  \BibitemOpen
  \bibfield  {author} {\bibinfo {author} {\bibfnamefont {A.}~\bibnamefont
  {Boyarsky}}, \bibinfo {author} {\bibfnamefont {J.}~\bibnamefont {Franse}},
  \bibinfo {author} {\bibfnamefont {D.}~\bibnamefont {Iakubovskyi}}, \ and\
  \bibinfo {author} {\bibfnamefont {O.}~\bibnamefont {Ruchayskiy}},\ }\href
  {\doibase 10.1103/PhysRevLett.115.161301} {\bibfield  {journal} {\bibinfo
  {journal} {Phys. Rev. Lett.}\ }\textbf {\bibinfo {volume} {115}},\ \bibinfo
  {pages} {161301} (\bibinfo {year} {2015})}\BibitemShut {NoStop}%
\bibitem [{\citenamefont {Iakubovskyi}\ \emph {et~al.}(2015)\citenamefont
  {Iakubovskyi}, \citenamefont {Bulbul}, \citenamefont {Foster}, \citenamefont
  {Savchenko},\ and\ \citenamefont {Sadova}}]{iakubovskyi2015testing}%
  \BibitemOpen
  \bibfield  {author} {\bibinfo {author} {\bibfnamefont {D.}~\bibnamefont
  {Iakubovskyi}}, \bibinfo {author} {\bibfnamefont {E.}~\bibnamefont {Bulbul}},
  \bibinfo {author} {\bibfnamefont {A.~R.}\ \bibnamefont {Foster}}, \bibinfo
  {author} {\bibfnamefont {D.}~\bibnamefont {Savchenko}}, \ and\ \bibinfo
  {author} {\bibfnamefont {V.}~\bibnamefont {Sadova}},\ }\href@noop {}
  {\enquote {\bibinfo {title} {Testing the origin of ~3.55 kev line in
  individual galaxy clusters observed with xmm-newton},}\ } (\bibinfo {year}
  {2015}),\ \Eprint {http://arxiv.org/abs/1508.05186} {arXiv:1508.05186
  [astro-ph.HE]} \BibitemShut {NoStop}%
\bibitem [{\citenamefont {Franse}\ \emph {et~al.}(2016)\citenamefont {Franse},
  \citenamefont {Bulbul}, \citenamefont {Foster}, \citenamefont {Boyarsky},
  \citenamefont {Markevitch}, \citenamefont {Bautz}, \citenamefont
  {Iakubovskyi}, \citenamefont {Loewenstein}, \citenamefont {McDonald},
  \citenamefont {Miller}, \citenamefont {Randall}, \citenamefont {Ruchayskiy},\
  and\ \citenamefont {Smith}}]{Franse_2016}%
  \BibitemOpen
  \bibfield  {author} {\bibinfo {author} {\bibfnamefont {J.}~\bibnamefont
  {Franse}}, \bibinfo {author} {\bibfnamefont {E.}~\bibnamefont {Bulbul}},
  \bibinfo {author} {\bibfnamefont {A.}~\bibnamefont {Foster}}, \bibinfo
  {author} {\bibfnamefont {A.}~\bibnamefont {Boyarsky}}, \bibinfo {author}
  {\bibfnamefont {M.}~\bibnamefont {Markevitch}}, \bibinfo {author}
  {\bibfnamefont {M.}~\bibnamefont {Bautz}}, \bibinfo {author} {\bibfnamefont
  {D.}~\bibnamefont {Iakubovskyi}}, \bibinfo {author} {\bibfnamefont
  {M.}~\bibnamefont {Loewenstein}}, \bibinfo {author} {\bibfnamefont
  {M.}~\bibnamefont {McDonald}}, \bibinfo {author} {\bibfnamefont
  {E.}~\bibnamefont {Miller}}, \bibinfo {author} {\bibfnamefont {S.~W.}\
  \bibnamefont {Randall}}, \bibinfo {author} {\bibfnamefont {O.}~\bibnamefont
  {Ruchayskiy}}, \ and\ \bibinfo {author} {\bibfnamefont {R.~K.}\ \bibnamefont
  {Smith}},\ }\href {\doibase 10.3847/0004-637x/829/2/124} {\bibfield
  {journal} {\bibinfo  {journal} {The Astrophysical Journal}\ }\textbf
  {\bibinfo {volume} {829}},\ \bibinfo {pages} {124} (\bibinfo {year}
  {2016})}\BibitemShut {NoStop}%
\bibitem [{\citenamefont {Cappelluti}\ \emph {et~al.}(2018)\citenamefont
  {Cappelluti}, \citenamefont {Bulbul}, \citenamefont {Foster}, \citenamefont
  {Natarajan}, \citenamefont {Urry}, \citenamefont {Bautz}, \citenamefont
  {Civano}, \citenamefont {Miller},\ and\ \citenamefont
  {Smith}}]{Cappelluti_2018}%
  \BibitemOpen
  \bibfield  {author} {\bibinfo {author} {\bibfnamefont {N.}~\bibnamefont
  {Cappelluti}}, \bibinfo {author} {\bibfnamefont {E.}~\bibnamefont {Bulbul}},
  \bibinfo {author} {\bibfnamefont {A.}~\bibnamefont {Foster}}, \bibinfo
  {author} {\bibfnamefont {P.}~\bibnamefont {Natarajan}}, \bibinfo {author}
  {\bibfnamefont {M.~C.}\ \bibnamefont {Urry}}, \bibinfo {author}
  {\bibfnamefont {M.~W.}\ \bibnamefont {Bautz}}, \bibinfo {author}
  {\bibfnamefont {F.}~\bibnamefont {Civano}}, \bibinfo {author} {\bibfnamefont
  {E.}~\bibnamefont {Miller}}, \ and\ \bibinfo {author} {\bibfnamefont {R.~K.}\
  \bibnamefont {Smith}},\ }\href {\doibase 10.3847/1538-4357/aaaa68} {\bibfield
   {journal} {\bibinfo  {journal} {The Astrophysical Journal}\ }\textbf
  {\bibinfo {volume} {854}},\ \bibinfo {pages} {179} (\bibinfo {year}
  {2018})}\BibitemShut {NoStop}%
\bibitem [{\citenamefont {Neronov}\ \emph {et~al.}(2016)\citenamefont
  {Neronov}, \citenamefont {Malyshev},\ and\ \citenamefont
  {Eckert}}]{Neronov_2016}%
  \BibitemOpen
  \bibfield  {author} {\bibinfo {author} {\bibfnamefont {A.}~\bibnamefont
  {Neronov}}, \bibinfo {author} {\bibfnamefont {D.}~\bibnamefont {Malyshev}}, \
  and\ \bibinfo {author} {\bibfnamefont {D.}~\bibnamefont {Eckert}},\ }\href
  {\doibase 10.1103/physrevd.94.123504} {\bibfield  {journal} {\bibinfo
  {journal} {Physical Review D}\ }\textbf {\bibinfo {volume} {94}} (\bibinfo
  {year} {2016}),\ 10.1103/physrevd.94.123504}\BibitemShut {NoStop}%
\bibitem [{\citenamefont {Urban}\ \emph {et~al.}(2014)\citenamefont {Urban},
  \citenamefont {Werner}, \citenamefont {Allen}, \citenamefont {Simionescu},
  \citenamefont {Kaastra},\ and\ \citenamefont {Strigari}}]{urban2014suzaku}%
  \BibitemOpen
  \bibfield  {author} {\bibinfo {author} {\bibfnamefont {O.}~\bibnamefont
  {Urban}}, \bibinfo {author} {\bibfnamefont {N.}~\bibnamefont {Werner}},
  \bibinfo {author} {\bibfnamefont {S.~W.}\ \bibnamefont {Allen}}, \bibinfo
  {author} {\bibfnamefont {A.}~\bibnamefont {Simionescu}}, \bibinfo {author}
  {\bibfnamefont {J.~S.}\ \bibnamefont {Kaastra}}, \ and\ \bibinfo {author}
  {\bibfnamefont {L.~E.}\ \bibnamefont {Strigari}},\ }\href@noop {} {\enquote
  {\bibinfo {title} {A suzaku search for dark matter emission lines in the
  x-ray brightest galaxy clusters},}\ } (\bibinfo {year} {2014}),\ \Eprint
  {http://arxiv.org/abs/1411.0050} {arXiv:1411.0050 [astro-ph.CO]} \BibitemShut
  {NoStop}%
\bibitem [{\citenamefont {Tamura}\ \emph {et~al.}(2015)\citenamefont {Tamura},
  \citenamefont {Iizuka}, \citenamefont {Maeda}, \citenamefont {Mitsuda},\ and\
  \citenamefont {Yamasaki}}]{Tamura_2015}%
  \BibitemOpen
  \bibfield  {author} {\bibinfo {author} {\bibfnamefont {T.}~\bibnamefont
  {Tamura}}, \bibinfo {author} {\bibfnamefont {R.}~\bibnamefont {Iizuka}},
  \bibinfo {author} {\bibfnamefont {Y.}~\bibnamefont {Maeda}}, \bibinfo
  {author} {\bibfnamefont {K.}~\bibnamefont {Mitsuda}}, \ and\ \bibinfo
  {author} {\bibfnamefont {N.~Y.}\ \bibnamefont {Yamasaki}},\ }\href {\doibase
  10.1093/pasj/psu156} {\bibfield  {journal} {\bibinfo  {journal} {Publications
  of the Astronomical Society of Japan}\ }\textbf {\bibinfo {volume} {67}}
  (\bibinfo {year} {2015}),\ 10.1093/pasj/psu156}\BibitemShut {NoStop}%
\bibitem [{\citenamefont {Riemer-S{\o}rensen}\ \emph
  {et~al.}(2015)\citenamefont {Riemer-S{\o}rensen}, \citenamefont {Wik},
  \citenamefont {Madejski}, \citenamefont {Molendi}, \citenamefont
  {Gastaldello}, \citenamefont {Harrison}, \citenamefont {Craig}, \citenamefont
  {Hailey}, \citenamefont {Boggs}, \citenamefont {Christensen}, \citenamefont
  {Stern}, \citenamefont {Zhang},\ and\ \citenamefont
  {Hornstrup}}]{Riemer_S_rensen_2015}%
  \BibitemOpen
  \bibfield  {author} {\bibinfo {author} {\bibfnamefont {S.}~\bibnamefont
  {Riemer-S{\o}rensen}}, \bibinfo {author} {\bibfnamefont {D.}~\bibnamefont
  {Wik}}, \bibinfo {author} {\bibfnamefont {G.}~\bibnamefont {Madejski}},
  \bibinfo {author} {\bibfnamefont {S.}~\bibnamefont {Molendi}}, \bibinfo
  {author} {\bibfnamefont {F.}~\bibnamefont {Gastaldello}}, \bibinfo {author}
  {\bibfnamefont {F.~A.}\ \bibnamefont {Harrison}}, \bibinfo {author}
  {\bibfnamefont {W.~W.}\ \bibnamefont {Craig}}, \bibinfo {author}
  {\bibfnamefont {C.~J.}\ \bibnamefont {Hailey}}, \bibinfo {author}
  {\bibfnamefont {S.~E.}\ \bibnamefont {Boggs}}, \bibinfo {author}
  {\bibfnamefont {F.~E.}\ \bibnamefont {Christensen}}, \bibinfo {author}
  {\bibfnamefont {D.}~\bibnamefont {Stern}}, \bibinfo {author} {\bibfnamefont
  {W.~W.}\ \bibnamefont {Zhang}}, \ and\ \bibinfo {author} {\bibfnamefont
  {A.}~\bibnamefont {Hornstrup}},\ }\href {\doibase 10.1088/0004-637x/810/1/48}
  {\bibfield  {journal} {\bibinfo  {journal} {The Astrophysical Journal}\
  }\textbf {\bibinfo {volume} {810}},\ \bibinfo {pages} {48} (\bibinfo {year}
  {2015})}\BibitemShut {NoStop}%
\bibitem [{\citenamefont {Malyshev}\ \emph {et~al.}(2014)\citenamefont
  {Malyshev}, \citenamefont {Neronov},\ and\ \citenamefont
  {Eckert}}]{PhysRevD.90.103506}%
  \BibitemOpen
  \bibfield  {author} {\bibinfo {author} {\bibfnamefont {D.}~\bibnamefont
  {Malyshev}}, \bibinfo {author} {\bibfnamefont {A.}~\bibnamefont {Neronov}}, \
  and\ \bibinfo {author} {\bibfnamefont {D.}~\bibnamefont {Eckert}},\ }\href
  {\doibase 10.1103/PhysRevD.90.103506} {\bibfield  {journal} {\bibinfo
  {journal} {Phys. Rev. D}\ }\textbf {\bibinfo {volume} {90}},\ \bibinfo
  {pages} {103506} (\bibinfo {year} {2014})}\BibitemShut {NoStop}%
\bibitem [{\citenamefont {Anderson}\ \emph {et~al.}(2015)\citenamefont
  {Anderson}, \citenamefont {Churazov},\ and\ \citenamefont
  {Bregman}}]{Anderson_2015}%
  \BibitemOpen
  \bibfield  {author} {\bibinfo {author} {\bibfnamefont {M.~E.}\ \bibnamefont
  {Anderson}}, \bibinfo {author} {\bibfnamefont {E.}~\bibnamefont {Churazov}},
  \ and\ \bibinfo {author} {\bibfnamefont {J.~N.}\ \bibnamefont {Bregman}},\
  }\href {\doibase 10.1093/mnras/stv1559} {\bibfield  {journal} {\bibinfo
  {journal} {Monthly Notices of the Royal Astronomical Society}\ }\textbf
  {\bibinfo {volume} {452}},\ \bibinfo {pages} {3905–3923} (\bibinfo {year}
  {2015})}\BibitemShut {NoStop}%
\bibitem [{\citenamefont {Foster}\ \emph {et~al.}(2021)\citenamefont {Foster},
  \citenamefont {Kongsore}, \citenamefont {Dessert}, \citenamefont {Park},
  \citenamefont {Rodd}, \citenamefont {Cranmer},\ and\ \citenamefont
  {Safdi}}]{foster2021deep}%
  \BibitemOpen
  \bibfield  {author} {\bibinfo {author} {\bibfnamefont {J.~W.}\ \bibnamefont
  {Foster}}, \bibinfo {author} {\bibfnamefont {M.}~\bibnamefont {Kongsore}},
  \bibinfo {author} {\bibfnamefont {C.}~\bibnamefont {Dessert}}, \bibinfo
  {author} {\bibfnamefont {Y.}~\bibnamefont {Park}}, \bibinfo {author}
  {\bibfnamefont {N.~L.}\ \bibnamefont {Rodd}}, \bibinfo {author}
  {\bibfnamefont {K.}~\bibnamefont {Cranmer}}, \ and\ \bibinfo {author}
  {\bibfnamefont {B.~R.}\ \bibnamefont {Safdi}},\ }\href@noop {} {\enquote
  {\bibinfo {title} {A deep search for decaying dark matter with xmm-newton
  blank-sky observations},}\ } (\bibinfo {year} {2021}),\ \Eprint
  {http://arxiv.org/abs/2102.02207} {arXiv:2102.02207 [astro-ph.CO]}
  \BibitemShut {NoStop}%
\bibitem [{\citenamefont {Bhargava}\ \emph {et~al.}(2020)\citenamefont
  {Bhargava}, \citenamefont {Giles}, \citenamefont {Romer}, \citenamefont
  {Jeltema}, \citenamefont {Mayers}, \citenamefont {Bermeo}, \citenamefont
  {Hilton}, \citenamefont {Wilkinson}, \citenamefont {Vergara}, \citenamefont
  {Collins},\ and\ \citenamefont {et~al.}}]{Bhargava_2020}%
  \BibitemOpen
  \bibfield  {author} {\bibinfo {author} {\bibfnamefont {S.}~\bibnamefont
  {Bhargava}}, \bibinfo {author} {\bibfnamefont {P.~A.}\ \bibnamefont {Giles}},
  \bibinfo {author} {\bibfnamefont {A.~K.}\ \bibnamefont {Romer}}, \bibinfo
  {author} {\bibfnamefont {T.}~\bibnamefont {Jeltema}}, \bibinfo {author}
  {\bibfnamefont {J.}~\bibnamefont {Mayers}}, \bibinfo {author} {\bibfnamefont
  {A.}~\bibnamefont {Bermeo}}, \bibinfo {author} {\bibfnamefont
  {M.}~\bibnamefont {Hilton}}, \bibinfo {author} {\bibfnamefont
  {R.}~\bibnamefont {Wilkinson}}, \bibinfo {author} {\bibfnamefont
  {C.}~\bibnamefont {Vergara}}, \bibinfo {author} {\bibfnamefont {C.~A.}\
  \bibnamefont {Collins}}, \ and\ \bibinfo {author} {\bibnamefont {et~al.}},\
  }\href {\doibase 10.1093/mnras/staa1829} {\bibfield  {journal} {\bibinfo
  {journal} {Monthly Notices of the Royal Astronomical Society}\ }\textbf
  {\bibinfo {volume} {497}},\ \bibinfo {pages} {656–671} (\bibinfo {year}
  {2020})}\BibitemShut {NoStop}%
\bibitem [{\citenamefont {Dessert}\ \emph {et~al.}(2020)\citenamefont
  {Dessert}, \citenamefont {Rodd},\ and\ \citenamefont {Safdi}}]{Dessert_2020}%
  \BibitemOpen
  \bibfield  {author} {\bibinfo {author} {\bibfnamefont {C.}~\bibnamefont
  {Dessert}}, \bibinfo {author} {\bibfnamefont {N.~L.}\ \bibnamefont {Rodd}}, \
  and\ \bibinfo {author} {\bibfnamefont {B.~R.}\ \bibnamefont {Safdi}},\ }\href
  {\doibase 10.1126/science.aaw3772} {\bibfield  {journal} {\bibinfo  {journal}
  {Science}\ }\textbf {\bibinfo {volume} {367}},\ \bibinfo {pages}
  {1465–1467} (\bibinfo {year} {2020})}\BibitemShut {NoStop}%
\bibitem [{\citenamefont {Abazajian}(2020)}]{Abazajian:2020unr}%
  \BibitemOpen
  \bibfield  {author} {\bibinfo {author} {\bibfnamefont {K.~N.}\ \bibnamefont
  {Abazajian}},\ }\href@noop {} {\  (\bibinfo {year} {2020})},\ \Eprint
  {http://arxiv.org/abs/2004.06170} {arXiv:2004.06170 [astro-ph.HE]}
  \BibitemShut {NoStop}%
\bibitem [{\citenamefont {Boyarsky}\ \emph {et~al.}(2020)\citenamefont
  {Boyarsky}, \citenamefont {Malyshev}, \citenamefont {Ruchayskiy},\ and\
  \citenamefont {Savchenko}}]{Boyarsky:2020hqb}%
  \BibitemOpen
  \bibfield  {author} {\bibinfo {author} {\bibfnamefont {A.}~\bibnamefont
  {Boyarsky}}, \bibinfo {author} {\bibfnamefont {D.}~\bibnamefont {Malyshev}},
  \bibinfo {author} {\bibfnamefont {O.}~\bibnamefont {Ruchayskiy}}, \ and\
  \bibinfo {author} {\bibfnamefont {D.}~\bibnamefont {Savchenko}},\ }\href@noop
  {} {\  (\bibinfo {year} {2020})},\ \Eprint {http://arxiv.org/abs/2004.06601}
  {arXiv:2004.06601 [astro-ph.CO]} \BibitemShut {NoStop}%
\bibitem [{\citenamefont {Arias}\ \emph {et~al.}(2012)\citenamefont {Arias},
  \citenamefont {Cadamuro}, \citenamefont {Goodsell}, \citenamefont {Jaeckel},
  \citenamefont {Redondo},\ and\ \citenamefont {Ringwald}}]{Arias_2012}%
  \BibitemOpen
  \bibfield  {author} {\bibinfo {author} {\bibfnamefont {P.}~\bibnamefont
  {Arias}}, \bibinfo {author} {\bibfnamefont {D.}~\bibnamefont {Cadamuro}},
  \bibinfo {author} {\bibfnamefont {M.}~\bibnamefont {Goodsell}}, \bibinfo
  {author} {\bibfnamefont {J.}~\bibnamefont {Jaeckel}}, \bibinfo {author}
  {\bibfnamefont {J.}~\bibnamefont {Redondo}}, \ and\ \bibinfo {author}
  {\bibfnamefont {A.}~\bibnamefont {Ringwald}},\ }\href {\doibase
  10.1088/1475-7516/2012/06/013} {\bibfield  {journal} {\bibinfo  {journal}
  {Journal of Cosmology and Astroparticle Physics}\ }\textbf {\bibinfo {volume}
  {2012}},\ \bibinfo {pages} {013} (\bibinfo {year} {2012})}\BibitemShut
  {NoStop}%
\bibitem [{\citenamefont {Takahashi}\ \emph {et~al.}(2020)\citenamefont
  {Takahashi}, \citenamefont {Yamada},\ and\ \citenamefont
  {Yin}}]{Takahashi:2020bpq}%
  \BibitemOpen
  \bibfield  {author} {\bibinfo {author} {\bibfnamefont {F.}~\bibnamefont
  {Takahashi}}, \bibinfo {author} {\bibfnamefont {M.}~\bibnamefont {Yamada}}, \
  and\ \bibinfo {author} {\bibfnamefont {W.}~\bibnamefont {Yin}},\ }\href
  {\doibase 10.1103/PhysRevLett.125.161801} {\bibfield  {journal} {\bibinfo
  {journal} {Phys. Rev. Lett.}\ }\textbf {\bibinfo {volume} {125}},\ \bibinfo
  {pages} {161801} (\bibinfo {year} {2020})},\ \Eprint
  {http://arxiv.org/abs/2006.10035} {arXiv:2006.10035 [hep-ph]} \BibitemShut
  {NoStop}%
\bibitem [{\citenamefont {Irastorza}\ and\ \citenamefont
  {Redondo}(2018)}]{Irastorza:2018dyq}%
  \BibitemOpen
  \bibfield  {author} {\bibinfo {author} {\bibfnamefont {I.~G.}\ \bibnamefont
  {Irastorza}}\ and\ \bibinfo {author} {\bibfnamefont {J.}~\bibnamefont
  {Redondo}},\ }\href {\doibase 10.1016/j.ppnp.2018.05.003} {\bibfield
  {journal} {\bibinfo  {journal} {Prog. Part. Nucl. Phys.}\ }\textbf {\bibinfo
  {volume} {102}},\ \bibinfo {pages} {89} (\bibinfo {year} {2018})},\ \Eprint
  {http://arxiv.org/abs/1801.08127} {arXiv:1801.08127 [hep-ph]} \BibitemShut
  {NoStop}%
\bibitem [{\citenamefont {Chaubey}\ \emph {et~al.}(2020)\citenamefont
  {Chaubey}, \citenamefont {Jaiswal},\ and\ \citenamefont
  {Ganguly}}]{Chaubey_2020}%
  \BibitemOpen
  \bibfield  {author} {\bibinfo {author} {\bibfnamefont {A.}~\bibnamefont
  {Chaubey}}, \bibinfo {author} {\bibfnamefont {M.~K.}\ \bibnamefont
  {Jaiswal}}, \ and\ \bibinfo {author} {\bibfnamefont {A.~K.}\ \bibnamefont
  {Ganguly}},\ }\href {\doibase 10.1103/physrevd.102.123029} {\bibfield
  {journal} {\bibinfo  {journal} {Physical Review D}\ }\textbf {\bibinfo
  {volume} {102}} (\bibinfo {year} {2020}),\
  10.1103/physrevd.102.123029}\BibitemShut {NoStop}%
\bibitem [{\citenamefont {Aprile}\ \emph {et~al.}(2020)\citenamefont {Aprile},
  \citenamefont {Aalbers}, \citenamefont {Agostini}, \citenamefont {Alfonsi},
  \citenamefont {Althueser}, \citenamefont {Amaro}, \citenamefont {Antochi},
  \citenamefont {Angelino}, \citenamefont {Angevaare}, \citenamefont
  {Arneodo},\ and\ \citenamefont {et~al.}}]{Aprile_2020}%
  \BibitemOpen
  \bibfield  {author} {\bibinfo {author} {\bibfnamefont {E.}~\bibnamefont
  {Aprile}}, \bibinfo {author} {\bibfnamefont {J.}~\bibnamefont {Aalbers}},
  \bibinfo {author} {\bibfnamefont {F.}~\bibnamefont {Agostini}}, \bibinfo
  {author} {\bibfnamefont {M.}~\bibnamefont {Alfonsi}}, \bibinfo {author}
  {\bibfnamefont {L.}~\bibnamefont {Althueser}}, \bibinfo {author}
  {\bibfnamefont {F.}~\bibnamefont {Amaro}}, \bibinfo {author} {\bibfnamefont
  {V.}~\bibnamefont {Antochi}}, \bibinfo {author} {\bibfnamefont
  {E.}~\bibnamefont {Angelino}}, \bibinfo {author} {\bibfnamefont
  {J.}~\bibnamefont {Angevaare}}, \bibinfo {author} {\bibfnamefont
  {F.}~\bibnamefont {Arneodo}}, \ and\ \bibinfo {author} {\bibnamefont
  {et~al.}},\ }\href {\doibase 10.1103/physrevd.102.072004} {\bibfield
  {journal} {\bibinfo  {journal} {Physical Review D}\ }\textbf {\bibinfo
  {volume} {102}} (\bibinfo {year} {2020}),\
  10.1103/physrevd.102.072004}\BibitemShut {NoStop}%
\bibitem [{\citenamefont {Nakayama}\ \emph {et~al.}(2014)\citenamefont
  {Nakayama}, \citenamefont {Takahashi},\ and\ \citenamefont
  {Yanagida}}]{Nakayama_2014}%
  \BibitemOpen
  \bibfield  {author} {\bibinfo {author} {\bibfnamefont {K.}~\bibnamefont
  {Nakayama}}, \bibinfo {author} {\bibfnamefont {F.}~\bibnamefont {Takahashi}},
  \ and\ \bibinfo {author} {\bibfnamefont {T.~T.}\ \bibnamefont {Yanagida}},\
  }\href {\doibase 10.1016/j.physletb.2014.05.035} {\bibfield  {journal}
  {\bibinfo  {journal} {Physics Letters B}\ }\textbf {\bibinfo {volume}
  {734}},\ \bibinfo {pages} {178–182} (\bibinfo {year} {2014})}\BibitemShut
  {NoStop}%
\bibitem [{\citenamefont {Pospelov}\ \emph {et~al.}(2008)\citenamefont
  {Pospelov}, \citenamefont {Ritz},\ and\ \citenamefont
  {Voloshin}}]{Pospelov_2008}%
  \BibitemOpen
  \bibfield  {author} {\bibinfo {author} {\bibfnamefont {M.}~\bibnamefont
  {Pospelov}}, \bibinfo {author} {\bibfnamefont {A.}~\bibnamefont {Ritz}}, \
  and\ \bibinfo {author} {\bibfnamefont {M.}~\bibnamefont {Voloshin}},\ }\href
  {\doibase 10.1103/physrevd.78.115012} {\bibfield  {journal} {\bibinfo
  {journal} {Physical Review D}\ }\textbf {\bibinfo {volume} {78}} (\bibinfo
  {year} {2008}),\ 10.1103/physrevd.78.115012}\BibitemShut {NoStop}%
\bibitem [{\citenamefont {Giannotti}\ \emph {et~al.}(2017)\citenamefont
  {Giannotti}, \citenamefont {Irastorza}, \citenamefont {Redondo},
  \citenamefont {Ringwald},\ and\ \citenamefont {Saikawa}}]{Giannotti:2017hny}%
  \BibitemOpen
  \bibfield  {author} {\bibinfo {author} {\bibfnamefont {M.}~\bibnamefont
  {Giannotti}}, \bibinfo {author} {\bibfnamefont {I.~G.}\ \bibnamefont
  {Irastorza}}, \bibinfo {author} {\bibfnamefont {J.}~\bibnamefont {Redondo}},
  \bibinfo {author} {\bibfnamefont {A.}~\bibnamefont {Ringwald}}, \ and\
  \bibinfo {author} {\bibfnamefont {K.}~\bibnamefont {Saikawa}},\ }\href
  {\doibase 10.1088/1475-7516/2017/10/010} {\bibfield  {journal} {\bibinfo
  {journal} {JCAP}\ }\textbf {\bibinfo {volume} {10}},\ \bibinfo {pages} {010}
  (\bibinfo {year} {2017})},\ \Eprint {http://arxiv.org/abs/1708.02111}
  {arXiv:1708.02111 [hep-ph]} \BibitemShut {NoStop}%
\bibitem [{\citenamefont {Miller~Bertolami}\ \emph {et~al.}(2014)\citenamefont
  {Miller~Bertolami}, \citenamefont {Melendez}, \citenamefont {Althaus},\ and\
  \citenamefont {Isern}}]{Bertolami:2014wua}%
  \BibitemOpen
  \bibfield  {author} {\bibinfo {author} {\bibfnamefont {M.~M.}\ \bibnamefont
  {Miller~Bertolami}}, \bibinfo {author} {\bibfnamefont {B.~E.}\ \bibnamefont
  {Melendez}}, \bibinfo {author} {\bibfnamefont {L.~G.}\ \bibnamefont
  {Althaus}}, \ and\ \bibinfo {author} {\bibfnamefont {J.}~\bibnamefont
  {Isern}},\ }\href {\doibase 10.1088/1475-7516/2014/10/069} {\bibfield
  {journal} {\bibinfo  {journal} {JCAP}\ }\textbf {\bibinfo {volume} {10}},\
  \bibinfo {pages} {069} (\bibinfo {year} {2014})},\ \Eprint
  {http://arxiv.org/abs/1406.7712} {arXiv:1406.7712 [hep-ph]} \BibitemShut
  {NoStop}%
\bibitem [{\citenamefont {Ayala}\ \emph {et~al.}(2014)\citenamefont {Ayala},
  \citenamefont {Dom\'\i{}nguez}, \citenamefont {Giannotti}, \citenamefont
  {Mirizzi},\ and\ \citenamefont {Straniero}}]{Ayala:2014pea}%
  \BibitemOpen
  \bibfield  {author} {\bibinfo {author} {\bibfnamefont {A.}~\bibnamefont
  {Ayala}}, \bibinfo {author} {\bibfnamefont {I.}~\bibnamefont
  {Dom\'\i{}nguez}}, \bibinfo {author} {\bibfnamefont {M.}~\bibnamefont
  {Giannotti}}, \bibinfo {author} {\bibfnamefont {A.}~\bibnamefont {Mirizzi}},
  \ and\ \bibinfo {author} {\bibfnamefont {O.}~\bibnamefont {Straniero}},\
  }\href {\doibase 10.1103/PhysRevLett.113.191302} {\bibfield  {journal}
  {\bibinfo  {journal} {Phys. Rev. Lett.}\ }\textbf {\bibinfo {volume} {113}},\
  \bibinfo {pages} {191302} (\bibinfo {year} {2014})},\ \Eprint
  {http://arxiv.org/abs/1406.6053} {arXiv:1406.6053 [astro-ph.SR]} \BibitemShut
  {NoStop}%
\bibitem [{\citenamefont {Corsico}\ \emph
  {et~al.}(2012{\natexlab{a}})\citenamefont {Corsico}, \citenamefont {Althaus},
  \citenamefont {Bertolami}, \citenamefont {Romero}, \citenamefont
  {Garcia-Berro}, \citenamefont {Isern},\ and\ \citenamefont
  {Kepler}}]{Corsico:2012ki}%
  \BibitemOpen
  \bibfield  {author} {\bibinfo {author} {\bibfnamefont {A.~H.}\ \bibnamefont
  {Corsico}}, \bibinfo {author} {\bibfnamefont {L.~G.}\ \bibnamefont
  {Althaus}}, \bibinfo {author} {\bibfnamefont {M.~M.~M.}\ \bibnamefont
  {Bertolami}}, \bibinfo {author} {\bibfnamefont {A.~D.}\ \bibnamefont
  {Romero}}, \bibinfo {author} {\bibfnamefont {E.}~\bibnamefont
  {Garcia-Berro}}, \bibinfo {author} {\bibfnamefont {J.}~\bibnamefont {Isern}},
  \ and\ \bibinfo {author} {\bibfnamefont {S.~O.}\ \bibnamefont {Kepler}},\
  }\href {\doibase 10.1111/j.1365-2966.2012.21401.x} {\bibfield  {journal}
  {\bibinfo  {journal} {Mon. Not. Roy. Astron. Soc.}\ }\textbf {\bibinfo
  {volume} {424}},\ \bibinfo {pages} {2792} (\bibinfo {year}
  {2012}{\natexlab{a}})},\ \Eprint {http://arxiv.org/abs/1205.6180}
  {arXiv:1205.6180 [astro-ph.SR]} \BibitemShut {NoStop}%
\bibitem [{\citenamefont {Corsico}\ \emph
  {et~al.}(2012{\natexlab{b}})\citenamefont {Corsico}, \citenamefont {Althaus},
  \citenamefont {Romero}, \citenamefont {Mukadam}, \citenamefont
  {Garcia-Berro}, \citenamefont {Isern}, \citenamefont {Kepler},\ and\
  \citenamefont {Corti}}]{Corsico:2012sh}%
  \BibitemOpen
  \bibfield  {author} {\bibinfo {author} {\bibfnamefont {A.~H.}\ \bibnamefont
  {Corsico}}, \bibinfo {author} {\bibfnamefont {L.~G.}\ \bibnamefont
  {Althaus}}, \bibinfo {author} {\bibfnamefont {A.~D.}\ \bibnamefont {Romero}},
  \bibinfo {author} {\bibfnamefont {A.~S.}\ \bibnamefont {Mukadam}}, \bibinfo
  {author} {\bibfnamefont {E.}~\bibnamefont {Garcia-Berro}}, \bibinfo {author}
  {\bibfnamefont {J.}~\bibnamefont {Isern}}, \bibinfo {author} {\bibfnamefont
  {S.~O.}\ \bibnamefont {Kepler}}, \ and\ \bibinfo {author} {\bibfnamefont
  {M.~A.}\ \bibnamefont {Corti}},\ }\href {\doibase
  10.1088/1475-7516/2012/12/010} {\bibfield  {journal} {\bibinfo  {journal}
  {JCAP}\ }\textbf {\bibinfo {volume} {12}},\ \bibinfo {pages} {010} (\bibinfo
  {year} {2012}{\natexlab{b}})},\ \Eprint {http://arxiv.org/abs/1211.3389}
  {arXiv:1211.3389 [astro-ph.SR]} \BibitemShut {NoStop}%
\bibitem [{\citenamefont {Giannotti}\ \emph {et~al.}(2016)\citenamefont
  {Giannotti}, \citenamefont {Irastorza}, \citenamefont {Redondo},\ and\
  \citenamefont {Ringwald}}]{Giannotti:2015kwo}%
  \BibitemOpen
  \bibfield  {author} {\bibinfo {author} {\bibfnamefont {M.}~\bibnamefont
  {Giannotti}}, \bibinfo {author} {\bibfnamefont {I.}~\bibnamefont
  {Irastorza}}, \bibinfo {author} {\bibfnamefont {J.}~\bibnamefont {Redondo}},
  \ and\ \bibinfo {author} {\bibfnamefont {A.}~\bibnamefont {Ringwald}},\
  }\href {\doibase 10.1088/1475-7516/2016/05/057} {\bibfield  {journal}
  {\bibinfo  {journal} {JCAP}\ }\textbf {\bibinfo {volume} {05}},\ \bibinfo
  {pages} {057} (\bibinfo {year} {2016})},\ \Eprint
  {http://arxiv.org/abs/1512.08108} {arXiv:1512.08108 [astro-ph.HE]}
  \BibitemShut {NoStop}%
\bibitem [{\citenamefont {Navarro}\ \emph
  {et~al.}(1997{\natexlab{a}})\citenamefont {Navarro}, \citenamefont {Frenk},\
  and\ \citenamefont {White}}]{Navarro_1997}%
  \BibitemOpen
  \bibfield  {author} {\bibinfo {author} {\bibfnamefont {J.~F.}\ \bibnamefont
  {Navarro}}, \bibinfo {author} {\bibfnamefont {C.~S.}\ \bibnamefont {Frenk}},
  \ and\ \bibinfo {author} {\bibfnamefont {S.~D.~M.}\ \bibnamefont {White}},\
  }\href {\doibase 10.1086/304888} {\bibfield  {journal} {\bibinfo  {journal}
  {The Astrophysical Journal}\ }\textbf {\bibinfo {volume} {490}},\ \bibinfo
  {pages} {493–508} (\bibinfo {year} {1997}{\natexlab{a}})}\BibitemShut
  {NoStop}%
\bibitem [{\citenamefont {Merloni}\ \emph {et~al.}(2012)\citenamefont
  {Merloni}, \citenamefont {Predehl}, \citenamefont {Becker}, \citenamefont
  {Böhringer}, \citenamefont {Boller}, \citenamefont {Brunner}, \citenamefont
  {Brusa}, \citenamefont {Dennerl}, \citenamefont {Freyberg}, \citenamefont
  {Friedrich}, \citenamefont {Georgakakis}, \citenamefont {Haberl},
  \citenamefont {Hasinger}, \citenamefont {Meidinger}, \citenamefont {Mohr},
  \citenamefont {Nandra}, \citenamefont {Rau}, \citenamefont {Reiprich},
  \citenamefont {Robrade}, \citenamefont {Salvato}, \citenamefont {Santangelo},
  \citenamefont {Sasaki}, \citenamefont {Schwope}, \citenamefont {Wilms},\ and\
  \citenamefont {the German~eROSITA Consortium}}]{merloni2012erosita}%
  \BibitemOpen
  \bibfield  {author} {\bibinfo {author} {\bibfnamefont {A.}~\bibnamefont
  {Merloni}}, \bibinfo {author} {\bibfnamefont {P.}~\bibnamefont {Predehl}},
  \bibinfo {author} {\bibfnamefont {W.}~\bibnamefont {Becker}}, \bibinfo
  {author} {\bibfnamefont {H.}~\bibnamefont {Böhringer}}, \bibinfo {author}
  {\bibfnamefont {T.}~\bibnamefont {Boller}}, \bibinfo {author} {\bibfnamefont
  {H.}~\bibnamefont {Brunner}}, \bibinfo {author} {\bibfnamefont
  {M.}~\bibnamefont {Brusa}}, \bibinfo {author} {\bibfnamefont
  {K.}~\bibnamefont {Dennerl}}, \bibinfo {author} {\bibfnamefont
  {M.}~\bibnamefont {Freyberg}}, \bibinfo {author} {\bibfnamefont
  {P.}~\bibnamefont {Friedrich}}, \bibinfo {author} {\bibfnamefont
  {A.}~\bibnamefont {Georgakakis}}, \bibinfo {author} {\bibfnamefont
  {F.}~\bibnamefont {Haberl}}, \bibinfo {author} {\bibfnamefont
  {G.}~\bibnamefont {Hasinger}}, \bibinfo {author} {\bibfnamefont
  {N.}~\bibnamefont {Meidinger}}, \bibinfo {author} {\bibfnamefont
  {J.}~\bibnamefont {Mohr}}, \bibinfo {author} {\bibfnamefont {K.}~\bibnamefont
  {Nandra}}, \bibinfo {author} {\bibfnamefont {A.}~\bibnamefont {Rau}},
  \bibinfo {author} {\bibfnamefont {T.~H.}\ \bibnamefont {Reiprich}}, \bibinfo
  {author} {\bibfnamefont {J.}~\bibnamefont {Robrade}}, \bibinfo {author}
  {\bibfnamefont {M.}~\bibnamefont {Salvato}}, \bibinfo {author} {\bibfnamefont
  {A.}~\bibnamefont {Santangelo}}, \bibinfo {author} {\bibfnamefont
  {M.}~\bibnamefont {Sasaki}}, \bibinfo {author} {\bibfnamefont
  {A.}~\bibnamefont {Schwope}}, \bibinfo {author} {\bibfnamefont
  {J.}~\bibnamefont {Wilms}}, \ and\ \bibinfo {author} {\bibnamefont {the
  German~eROSITA Consortium}},\ }\href@noop {} {\enquote {\bibinfo {title}
  {erosita science book: Mapping the structure of the energetic universe},}\ }
  (\bibinfo {year} {2012}),\ \Eprint {http://arxiv.org/abs/1209.3114}
  {arXiv:1209.3114 [astro-ph.HE]} \BibitemShut {NoStop}%
\bibitem [{\citenamefont {Navarro}\ \emph
  {et~al.}(1997{\natexlab{b}})\citenamefont {Navarro}, \citenamefont {Frenk},\
  and\ \citenamefont {White}}]{Navarro:1996gj}%
  \BibitemOpen
  \bibfield  {author} {\bibinfo {author} {\bibfnamefont {J.~F.}\ \bibnamefont
  {Navarro}}, \bibinfo {author} {\bibfnamefont {C.~S.}\ \bibnamefont {Frenk}},
  \ and\ \bibinfo {author} {\bibfnamefont {S.~D.~M.}\ \bibnamefont {White}},\
  }\href {\doibase 10.1086/304888} {\bibfield  {journal} {\bibinfo  {journal}
  {Astrophys. J.}\ }\textbf {\bibinfo {volume} {490}},\ \bibinfo {pages} {493}
  (\bibinfo {year} {1997}{\natexlab{b}})},\ \Eprint
  {http://arxiv.org/abs/astro-ph/9611107} {arXiv:astro-ph/9611107 [astro-ph]}
  \BibitemShut {NoStop}%
CITATION = ASTRO-PH/9611107;
\bibitem [{\citenamefont {Read}\ \emph {et~al.}(2016)\citenamefont {Read},
  \citenamefont {Agertz},\ and\ \citenamefont
  {Collins}}]{10.1093/mnras/stw713}%
  \BibitemOpen
  \bibfield  {author} {\bibinfo {author} {\bibfnamefont {J.~I.}\ \bibnamefont
  {Read}}, \bibinfo {author} {\bibfnamefont {O.}~\bibnamefont {Agertz}}, \ and\
  \bibinfo {author} {\bibfnamefont {M.~L.~M.}\ \bibnamefont {Collins}},\ }\href
  {\doibase 10.1093/mnras/stw713} {\bibfield  {journal} {\bibinfo  {journal}
  {Monthly Notices of the Royal Astronomical Society}\ }\textbf {\bibinfo
  {volume} {459}},\ \bibinfo {pages} {2573} (\bibinfo {year} {2016})},\ \Eprint
  {http://arxiv.org/abs/https://academic.oup.com/mnras/article-pdf/459/3/2573/8105757/stw713.pdf}
  {https://academic.oup.com/mnras/article-pdf/459/3/2573/8105757/stw713.pdf}
  \BibitemShut {NoStop}%
\bibitem [{\citenamefont {Abazajian}\ \emph {et~al.}(2020)\citenamefont
  {Abazajian}, \citenamefont {Horiuchi}, \citenamefont {Kaplinghat},
  \citenamefont {Keeley},\ and\ \citenamefont {Macias}}]{PhysRevD.102.043012}%
  \BibitemOpen
  \bibfield  {author} {\bibinfo {author} {\bibfnamefont {K.~N.}\ \bibnamefont
  {Abazajian}}, \bibinfo {author} {\bibfnamefont {S.}~\bibnamefont {Horiuchi}},
  \bibinfo {author} {\bibfnamefont {M.}~\bibnamefont {Kaplinghat}}, \bibinfo
  {author} {\bibfnamefont {R.~E.}\ \bibnamefont {Keeley}}, \ and\ \bibinfo
  {author} {\bibfnamefont {O.}~\bibnamefont {Macias}},\ }\href {\doibase
  10.1103/PhysRevD.102.043012} {\bibfield  {journal} {\bibinfo  {journal}
  {Phys. Rev. D}\ }\textbf {\bibinfo {volume} {102}},\ \bibinfo {pages}
  {043012} (\bibinfo {year} {2020})}\BibitemShut {NoStop}%
\bibitem [{\citenamefont {Higaki}\ \emph {et~al.}(2014)\citenamefont {Higaki},
  \citenamefont {Jeong},\ and\ \citenamefont {Takahashi}}]{Higaki:2014zua}%
  \BibitemOpen
  \bibfield  {author} {\bibinfo {author} {\bibfnamefont {T.}~\bibnamefont
  {Higaki}}, \bibinfo {author} {\bibfnamefont {K.~S.}\ \bibnamefont {Jeong}}, \
  and\ \bibinfo {author} {\bibfnamefont {F.}~\bibnamefont {Takahashi}},\ }\href
  {\doibase 10.1016/j.physletb.2014.04.007} {\bibfield  {journal} {\bibinfo
  {journal} {Phys. Lett. B}\ }\textbf {\bibinfo {volume} {733}},\ \bibinfo
  {pages} {25} (\bibinfo {year} {2014})},\ \Eprint
  {http://arxiv.org/abs/1402.6965} {arXiv:1402.6965 [hep-ph]} \BibitemShut
  {NoStop}%
\bibitem [{\citenamefont {Gorski}\ \emph {et~al.}(2005)\citenamefont {Gorski},
  \citenamefont {Hivon}, \citenamefont {Banday}, \citenamefont {Wandelt},
  \citenamefont {Hansen}, \citenamefont {Reinecke},\ and\ \citenamefont
  {Bartelman}}]{Gorski:2004by}%
  \BibitemOpen
  \bibfield  {author} {\bibinfo {author} {\bibfnamefont {K.}~\bibnamefont
  {Gorski}}, \bibinfo {author} {\bibfnamefont {E.}~\bibnamefont {Hivon}},
  \bibinfo {author} {\bibfnamefont {A.}~\bibnamefont {Banday}}, \bibinfo
  {author} {\bibfnamefont {B.}~\bibnamefont {Wandelt}}, \bibinfo {author}
  {\bibfnamefont {F.}~\bibnamefont {Hansen}}, \bibinfo {author} {\bibfnamefont
  {M.}~\bibnamefont {Reinecke}}, \ and\ \bibinfo {author} {\bibfnamefont
  {M.}~\bibnamefont {Bartelman}},\ }\href {\doibase 10.1086/427976} {\bibfield
  {journal} {\bibinfo  {journal} {Astrophys. J.}\ }\textbf {\bibinfo {volume}
  {622}},\ \bibinfo {pages} {759} (\bibinfo {year} {2005})},\ \Eprint
  {http://arxiv.org/abs/astro-ph/0409513} {arXiv:astro-ph/0409513} \BibitemShut
  {NoStop}%
\bibitem [{\citenamefont {Zonca}\ \emph {et~al.}(2019)\citenamefont {Zonca},
  \citenamefont {Singer}, \citenamefont {Lenz}, \citenamefont {Reinecke},
  \citenamefont {Rosset}, \citenamefont {Hivon},\ and\ \citenamefont
  {Gorski}}]{Zonca2019}%
  \BibitemOpen
  \bibfield  {author} {\bibinfo {author} {\bibfnamefont {A.}~\bibnamefont
  {Zonca}}, \bibinfo {author} {\bibfnamefont {L.}~\bibnamefont {Singer}},
  \bibinfo {author} {\bibfnamefont {D.}~\bibnamefont {Lenz}}, \bibinfo {author}
  {\bibfnamefont {M.}~\bibnamefont {Reinecke}}, \bibinfo {author}
  {\bibfnamefont {C.}~\bibnamefont {Rosset}}, \bibinfo {author} {\bibfnamefont
  {E.}~\bibnamefont {Hivon}}, \ and\ \bibinfo {author} {\bibfnamefont
  {K.}~\bibnamefont {Gorski}},\ }\href {\doibase 10.21105/joss.01298}
  {\bibfield  {journal} {\bibinfo  {journal} {Journal of Open Source Software}\
  }\textbf {\bibinfo {volume} {4}},\ \bibinfo {pages} {1298} (\bibinfo {year}
  {2019})}\BibitemShut {NoStop}%
\bibitem [{\citenamefont {Lumb}\ \emph {et~al.}(2002)\citenamefont {Lumb},
  \citenamefont {Warwick}, \citenamefont {Page},\ and\ \citenamefont
  {De~Luca}}]{Lumb:2002sw}%
  \BibitemOpen
  \bibfield  {author} {\bibinfo {author} {\bibfnamefont {D.}~\bibnamefont
  {Lumb}}, \bibinfo {author} {\bibfnamefont {R.}~\bibnamefont {Warwick}},
  \bibinfo {author} {\bibfnamefont {M.}~\bibnamefont {Page}}, \ and\ \bibinfo
  {author} {\bibfnamefont {A.}~\bibnamefont {De~Luca}},\ }\href {\doibase
  10.1051/0004-6361:20020531} {\bibfield  {journal} {\bibinfo  {journal}
  {Astron. Astrophys.}\ }\textbf {\bibinfo {volume} {389}},\ \bibinfo {pages}
  {93} (\bibinfo {year} {2002})},\ \Eprint
  {http://arxiv.org/abs/astro-ph/0204147} {arXiv:astro-ph/0204147} \BibitemShut
  {NoStop}%
\bibitem [{\citenamefont {Predehl}\ \emph {et~al.}(2021)\citenamefont
  {Predehl}, \citenamefont {Andritschke}, \citenamefont {Arefiev},
  \citenamefont {Babyshkin}, \citenamefont {Batanov}, \citenamefont {Becker},
  \citenamefont {Böhringer}, \citenamefont {Bogomolov}, \citenamefont
  {Boller}, \citenamefont {Borm},\ and\ \citenamefont {et~al.}}]{Predehl_2021}%
  \BibitemOpen
  \bibfield  {author} {\bibinfo {author} {\bibfnamefont {P.}~\bibnamefont
  {Predehl}}, \bibinfo {author} {\bibfnamefont {R.}~\bibnamefont
  {Andritschke}}, \bibinfo {author} {\bibfnamefont {V.}~\bibnamefont
  {Arefiev}}, \bibinfo {author} {\bibfnamefont {V.}~\bibnamefont {Babyshkin}},
  \bibinfo {author} {\bibfnamefont {O.}~\bibnamefont {Batanov}}, \bibinfo
  {author} {\bibfnamefont {W.}~\bibnamefont {Becker}}, \bibinfo {author}
  {\bibfnamefont {H.}~\bibnamefont {Böhringer}}, \bibinfo {author}
  {\bibfnamefont {A.}~\bibnamefont {Bogomolov}}, \bibinfo {author}
  {\bibfnamefont {T.}~\bibnamefont {Boller}}, \bibinfo {author} {\bibfnamefont
  {K.}~\bibnamefont {Borm}}, \ and\ \bibinfo {author} {\bibnamefont {et~al.}},\
  }\href {\doibase 10.1051/0004-6361/202039313} {\bibfield  {journal} {\bibinfo
   {journal} {Astronomy \& Astrophysics}\ }\textbf {\bibinfo {volume} {647}},\
  \bibinfo {pages} {A1} (\bibinfo {year} {2021})}\BibitemShut {NoStop}%
\bibitem [{\citenamefont {Predehl}\ \emph {et~al.}(2020)\citenamefont
  {Predehl}, \citenamefont {Sunyaev}, \citenamefont {Becker}, \citenamefont
  {Brunner}, \citenamefont {Burenin}, \citenamefont {Bykov}, \citenamefont
  {Cherepashchuk}, \citenamefont {Chugai}, \citenamefont {Churazov},
  \citenamefont {Doroshenko},\ and\ \citenamefont {et~al.}}]{Predehl_2020}%
  \BibitemOpen
  \bibfield  {author} {\bibinfo {author} {\bibfnamefont {P.}~\bibnamefont
  {Predehl}}, \bibinfo {author} {\bibfnamefont {R.~A.}\ \bibnamefont
  {Sunyaev}}, \bibinfo {author} {\bibfnamefont {W.}~\bibnamefont {Becker}},
  \bibinfo {author} {\bibfnamefont {H.}~\bibnamefont {Brunner}}, \bibinfo
  {author} {\bibfnamefont {R.}~\bibnamefont {Burenin}}, \bibinfo {author}
  {\bibfnamefont {A.}~\bibnamefont {Bykov}}, \bibinfo {author} {\bibfnamefont
  {A.}~\bibnamefont {Cherepashchuk}}, \bibinfo {author} {\bibfnamefont
  {N.}~\bibnamefont {Chugai}}, \bibinfo {author} {\bibfnamefont
  {E.}~\bibnamefont {Churazov}}, \bibinfo {author} {\bibfnamefont
  {V.}~\bibnamefont {Doroshenko}}, \ and\ \bibinfo {author} {\bibnamefont
  {et~al.}},\ }\href {\doibase 10.1038/s41586-020-2979-0} {\bibfield  {journal}
  {\bibinfo  {journal} {Nature}\ }\textbf {\bibinfo {volume} {588}},\ \bibinfo
  {pages} {227–231} (\bibinfo {year} {2020})}\BibitemShut {NoStop}%
\bibitem [{\citenamefont {Horiuchi}\ \emph {et~al.}(2014)\citenamefont
  {Horiuchi}, \citenamefont {Humphrey}, \citenamefont {Onorbe}, \citenamefont
  {Abazajian}, \citenamefont {Kaplinghat},\ and\ \citenamefont
  {Garrison-Kimmel}}]{Horiuchi:2013noa}%
  \BibitemOpen
  \bibfield  {author} {\bibinfo {author} {\bibfnamefont {S.}~\bibnamefont
  {Horiuchi}}, \bibinfo {author} {\bibfnamefont {P.~J.}\ \bibnamefont
  {Humphrey}}, \bibinfo {author} {\bibfnamefont {J.}~\bibnamefont {Onorbe}},
  \bibinfo {author} {\bibfnamefont {K.~N.}\ \bibnamefont {Abazajian}}, \bibinfo
  {author} {\bibfnamefont {M.}~\bibnamefont {Kaplinghat}}, \ and\ \bibinfo
  {author} {\bibfnamefont {S.}~\bibnamefont {Garrison-Kimmel}},\ }\href
  {\doibase 10.1103/PhysRevD.89.025017} {\bibfield  {journal} {\bibinfo
  {journal} {Phys. Rev. D}\ }\textbf {\bibinfo {volume} {89}},\ \bibinfo
  {pages} {025017} (\bibinfo {year} {2014})},\ \Eprint
  {http://arxiv.org/abs/1311.0282} {arXiv:1311.0282 [astro-ph.CO]} \BibitemShut
  {NoStop}%
\bibitem [{\citenamefont {Ng}\ \emph {et~al.}(2015)\citenamefont {Ng},
  \citenamefont {Horiuchi}, \citenamefont {Gaskins}, \citenamefont {Smith},\
  and\ \citenamefont {Preece}}]{Ng:2015gfa}%
  \BibitemOpen
  \bibfield  {author} {\bibinfo {author} {\bibfnamefont {K.~C.~Y.}\
  \bibnamefont {Ng}}, \bibinfo {author} {\bibfnamefont {S.}~\bibnamefont
  {Horiuchi}}, \bibinfo {author} {\bibfnamefont {J.~M.}\ \bibnamefont
  {Gaskins}}, \bibinfo {author} {\bibfnamefont {M.}~\bibnamefont {Smith}}, \
  and\ \bibinfo {author} {\bibfnamefont {R.}~\bibnamefont {Preece}},\ }\href
  {\doibase 10.1103/PhysRevD.92.043503} {\bibfield  {journal} {\bibinfo
  {journal} {Phys. Rev. D}\ }\textbf {\bibinfo {volume} {92}},\ \bibinfo
  {pages} {043503} (\bibinfo {year} {2015})},\ \Eprint
  {http://arxiv.org/abs/1504.04027} {arXiv:1504.04027 [astro-ph.CO]}
  \BibitemShut {NoStop}%
\bibitem [{\citenamefont {Perez}\ \emph {et~al.}(2017)\citenamefont {Perez},
  \citenamefont {Ng}, \citenamefont {Beacom}, \citenamefont {Hersh},
  \citenamefont {Horiuchi},\ and\ \citenamefont {Krivonos}}]{Perez:2016tcq}%
  \BibitemOpen
  \bibfield  {author} {\bibinfo {author} {\bibfnamefont {K.}~\bibnamefont
  {Perez}}, \bibinfo {author} {\bibfnamefont {K.~C.~Y.}\ \bibnamefont {Ng}},
  \bibinfo {author} {\bibfnamefont {J.~F.}\ \bibnamefont {Beacom}}, \bibinfo
  {author} {\bibfnamefont {C.}~\bibnamefont {Hersh}}, \bibinfo {author}
  {\bibfnamefont {S.}~\bibnamefont {Horiuchi}}, \ and\ \bibinfo {author}
  {\bibfnamefont {R.}~\bibnamefont {Krivonos}},\ }\href {\doibase
  10.1103/PhysRevD.95.123002} {\bibfield  {journal} {\bibinfo  {journal} {Phys.
  Rev. D}\ }\textbf {\bibinfo {volume} {95}},\ \bibinfo {pages} {123002}
  (\bibinfo {year} {2017})},\ \Eprint {http://arxiv.org/abs/1609.00667}
  {arXiv:1609.00667 [astro-ph.HE]} \BibitemShut {NoStop}%
\bibitem [{\citenamefont {Ng}\ \emph {et~al.}(2019)\citenamefont {Ng},
  \citenamefont {Roach}, \citenamefont {Perez}, \citenamefont {Beacom},
  \citenamefont {Horiuchi}, \citenamefont {Krivonos},\ and\ \citenamefont
  {Wik}}]{Ng:2019gch}%
  \BibitemOpen
  \bibfield  {author} {\bibinfo {author} {\bibfnamefont {K.~C.~Y.}\
  \bibnamefont {Ng}}, \bibinfo {author} {\bibfnamefont {B.~M.}\ \bibnamefont
  {Roach}}, \bibinfo {author} {\bibfnamefont {K.}~\bibnamefont {Perez}},
  \bibinfo {author} {\bibfnamefont {J.~F.}\ \bibnamefont {Beacom}}, \bibinfo
  {author} {\bibfnamefont {S.}~\bibnamefont {Horiuchi}}, \bibinfo {author}
  {\bibfnamefont {R.}~\bibnamefont {Krivonos}}, \ and\ \bibinfo {author}
  {\bibfnamefont {D.~R.}\ \bibnamefont {Wik}},\ }\href {\doibase
  10.1103/PhysRevD.99.083005} {\bibfield  {journal} {\bibinfo  {journal} {Phys.
  Rev. D}\ }\textbf {\bibinfo {volume} {99}},\ \bibinfo {pages} {083005}
  (\bibinfo {year} {2019})},\ \Eprint {http://arxiv.org/abs/1901.01262}
  {arXiv:1901.01262 [astro-ph.HE]} \BibitemShut {NoStop}%
\bibitem [{\citenamefont {Abazajian}(2017)}]{Abazajian_2017}%
  \BibitemOpen
  \bibfield  {author} {\bibinfo {author} {\bibfnamefont {K.~N.}\ \bibnamefont
  {Abazajian}},\ }\href {\doibase 10.1016/j.physrep.2017.10.003} {\bibfield
  {journal} {\bibinfo  {journal} {Physics Reports}\ }\textbf {\bibinfo {volume}
  {711-712}},\ \bibinfo {pages} {1–28} (\bibinfo {year} {2017})}\BibitemShut
  {NoStop}%
\bibitem [{\citenamefont {Caputo}\ \emph {et~al.}(2020)\citenamefont {Caputo},
  \citenamefont {Regis},\ and\ \citenamefont {Taoso}}]{Caputo:2019djj}%
  \BibitemOpen
  \bibfield  {author} {\bibinfo {author} {\bibfnamefont {A.}~\bibnamefont
  {Caputo}}, \bibinfo {author} {\bibfnamefont {M.}~\bibnamefont {Regis}}, \
  and\ \bibinfo {author} {\bibfnamefont {M.}~\bibnamefont {Taoso}},\ }\href
  {\doibase 10.1088/1475-7516/2020/03/001} {\bibfield  {journal} {\bibinfo
  {journal} {JCAP}\ }\textbf {\bibinfo {volume} {03}},\ \bibinfo {pages} {001}
  (\bibinfo {year} {2020})},\ \Eprint {http://arxiv.org/abs/1911.09120}
  {arXiv:1911.09120 [astro-ph.CO]} \BibitemShut {NoStop}%
\bibitem [{\citenamefont {Roach}\ \emph {et~al.}(2020)\citenamefont {Roach},
  \citenamefont {Ng}, \citenamefont {Perez}, \citenamefont {Beacom},
  \citenamefont {Horiuchi}, \citenamefont {Krivonos},\ and\ \citenamefont
  {Wik}}]{Roach_2020}%
  \BibitemOpen
  \bibfield  {author} {\bibinfo {author} {\bibfnamefont {B.~M.}\ \bibnamefont
  {Roach}}, \bibinfo {author} {\bibfnamefont {K.~C.}\ \bibnamefont {Ng}},
  \bibinfo {author} {\bibfnamefont {K.}~\bibnamefont {Perez}}, \bibinfo
  {author} {\bibfnamefont {J.~F.}\ \bibnamefont {Beacom}}, \bibinfo {author}
  {\bibfnamefont {S.}~\bibnamefont {Horiuchi}}, \bibinfo {author}
  {\bibfnamefont {R.}~\bibnamefont {Krivonos}}, \ and\ \bibinfo {author}
  {\bibfnamefont {D.~R.}\ \bibnamefont {Wik}},\ }\href {\doibase
  10.1103/physrevd.101.103011} {\bibfield  {journal} {\bibinfo  {journal}
  {Physical Review D}\ }\textbf {\bibinfo {volume} {101}} (\bibinfo {year}
  {2020}),\ 10.1103/physrevd.101.103011}\BibitemShut {NoStop}%
\bibitem [{\citenamefont {Serpico}\ and\ \citenamefont
  {Raffelt}(2005)}]{Serpico_2005}%
  \BibitemOpen
  \bibfield  {author} {\bibinfo {author} {\bibfnamefont {P.~D.}\ \bibnamefont
  {Serpico}}\ and\ \bibinfo {author} {\bibfnamefont {G.~G.}\ \bibnamefont
  {Raffelt}},\ }\href {\doibase 10.1103/physrevd.71.127301} {\bibfield
  {journal} {\bibinfo  {journal} {Physical Review D}\ }\textbf {\bibinfo
  {volume} {71}} (\bibinfo {year} {2005}),\
  10.1103/physrevd.71.127301}\BibitemShut {NoStop}%
\bibitem [{\citenamefont {Cherry}\ and\ \citenamefont
  {Horiuchi}(2017)}]{Cherry_2017}%
  \BibitemOpen
  \bibfield  {author} {\bibinfo {author} {\bibfnamefont {J.~F.}\ \bibnamefont
  {Cherry}}\ and\ \bibinfo {author} {\bibfnamefont {S.}~\bibnamefont
  {Horiuchi}},\ }\href {\doibase 10.1103/physrevd.95.083015} {\bibfield
  {journal} {\bibinfo  {journal} {Physical Review D}\ }\textbf {\bibinfo
  {volume} {95}} (\bibinfo {year} {2017}),\
  10.1103/physrevd.95.083015}\BibitemShut {NoStop}%
\bibitem [{\citenamefont {Barinov}\ \emph {et~al.}(2021)\citenamefont
  {Barinov}, \citenamefont {Burenin}, \citenamefont {Gorbunov},\ and\
  \citenamefont {Krivonos}}]{Barinov_2021}%
  \BibitemOpen
  \bibfield  {author} {\bibinfo {author} {\bibfnamefont {V.}~\bibnamefont
  {Barinov}}, \bibinfo {author} {\bibfnamefont {R.}~\bibnamefont {Burenin}},
  \bibinfo {author} {\bibfnamefont {D.}~\bibnamefont {Gorbunov}}, \ and\
  \bibinfo {author} {\bibfnamefont {R.}~\bibnamefont {Krivonos}},\ }\href
  {\doibase 10.1103/physrevd.103.063512} {\bibfield  {journal} {\bibinfo
  {journal} {Physical Review D}\ }\textbf {\bibinfo {volume} {103}} (\bibinfo
  {year} {2021}),\ 10.1103/physrevd.103.063512}\BibitemShut {NoStop}%
\bibitem [{\citenamefont {Neronov}\ and\ \citenamefont
  {Malyshev}(2016)}]{Neronov_2016_2}%
  \BibitemOpen
  \bibfield  {author} {\bibinfo {author} {\bibfnamefont {A.}~\bibnamefont
  {Neronov}}\ and\ \bibinfo {author} {\bibfnamefont {D.}~\bibnamefont
  {Malyshev}},\ }\href {\doibase 10.1103/physrevd.93.063518} {\bibfield
  {journal} {\bibinfo  {journal} {Physical Review D}\ }\textbf {\bibinfo
  {volume} {93}} (\bibinfo {year} {2016}),\
  10.1103/physrevd.93.063518}\BibitemShut {NoStop}%
\bibitem [{\citenamefont {Viaux}\ \emph {et~al.}(2013)\citenamefont {Viaux},
  \citenamefont {Catelan}, \citenamefont {Stetson}, \citenamefont {Raffelt},
  \citenamefont {Redondo}, \citenamefont {Valcarce},\ and\ \citenamefont
  {Weiss}}]{Viaux_2013}%
  \BibitemOpen
  \bibfield  {author} {\bibinfo {author} {\bibfnamefont {N.}~\bibnamefont
  {Viaux}}, \bibinfo {author} {\bibfnamefont {M.}~\bibnamefont {Catelan}},
  \bibinfo {author} {\bibfnamefont {P.~B.}\ \bibnamefont {Stetson}}, \bibinfo
  {author} {\bibfnamefont {G.~G.}\ \bibnamefont {Raffelt}}, \bibinfo {author}
  {\bibfnamefont {J.}~\bibnamefont {Redondo}}, \bibinfo {author} {\bibfnamefont
  {A.~A.~R.}\ \bibnamefont {Valcarce}}, \ and\ \bibinfo {author} {\bibfnamefont
  {A.}~\bibnamefont {Weiss}},\ }\href {\doibase 10.1103/physrevlett.111.231301}
  {\bibfield  {journal} {\bibinfo  {journal} {Physical Review Letters}\
  }\textbf {\bibinfo {volume} {111}} (\bibinfo {year} {2013}),\
  10.1103/physrevlett.111.231301}\BibitemShut {NoStop}%
\bibitem [{\citenamefont {Ando}\ \emph {et~al.}(2021)\citenamefont {Ando},
  \citenamefont {Barik}, \citenamefont {Feng}, \citenamefont {Finetti},
  \citenamefont {Chaves}, \citenamefont {Kanuri}, \citenamefont {Kleverlaan},
  \citenamefont {Ma}, \citenamefont {Serracapriola}, \citenamefont {Meinema},
  \citenamefont {Martinez}, \citenamefont {Ng}, \citenamefont {Peerbooms},
  \citenamefont {van Veen},\ and\ \citenamefont {Zimmer}}]{ando2021decaying}%
  \BibitemOpen
  \bibfield  {author} {\bibinfo {author} {\bibfnamefont {S.}~\bibnamefont
  {Ando}}, \bibinfo {author} {\bibfnamefont {S.~K.}\ \bibnamefont {Barik}},
  \bibinfo {author} {\bibfnamefont {Z.}~\bibnamefont {Feng}}, \bibinfo {author}
  {\bibfnamefont {M.}~\bibnamefont {Finetti}}, \bibinfo {author} {\bibfnamefont
  {A.~G.}\ \bibnamefont {Chaves}}, \bibinfo {author} {\bibfnamefont
  {S.}~\bibnamefont {Kanuri}}, \bibinfo {author} {\bibfnamefont
  {J.}~\bibnamefont {Kleverlaan}}, \bibinfo {author} {\bibfnamefont
  {Y.}~\bibnamefont {Ma}}, \bibinfo {author} {\bibfnamefont {N.~M.~D.}\
  \bibnamefont {Serracapriola}}, \bibinfo {author} {\bibfnamefont {M.~S.~P.}\
  \bibnamefont {Meinema}}, \bibinfo {author} {\bibfnamefont {I.~N.}\
  \bibnamefont {Martinez}}, \bibinfo {author} {\bibfnamefont {K.~C.~Y.}\
  \bibnamefont {Ng}}, \bibinfo {author} {\bibfnamefont {E.}~\bibnamefont
  {Peerbooms}}, \bibinfo {author} {\bibfnamefont {C.~A.}\ \bibnamefont {van
  Veen}}, \ and\ \bibinfo {author} {\bibfnamefont {F.}~\bibnamefont {Zimmer}},\
  }\href@noop {} {\enquote {\bibinfo {title} {Decaying dark matter in dwarf
  spheroidal galaxies: Prospects for x-ray and gamma-ray telescopes},}\ }
  (\bibinfo {year} {2021}),\ \Eprint {http://arxiv.org/abs/2103.13242}
  {arXiv:2103.13242 [astro-ph.HE]} \BibitemShut {NoStop}%
\bibitem [{\citenamefont {Speckhard}\ \emph {et~al.}(2016)\citenamefont
  {Speckhard}, \citenamefont {Ng}, \citenamefont {Beacom},\ and\ \citenamefont
  {Laha}}]{Speckhard:2015eva}%
  \BibitemOpen
  \bibfield  {author} {\bibinfo {author} {\bibfnamefont {E.~G.}\ \bibnamefont
  {Speckhard}}, \bibinfo {author} {\bibfnamefont {K.~C.~Y.}\ \bibnamefont
  {Ng}}, \bibinfo {author} {\bibfnamefont {J.~F.}\ \bibnamefont {Beacom}}, \
  and\ \bibinfo {author} {\bibfnamefont {R.}~\bibnamefont {Laha}},\ }\href
  {\doibase 10.1103/PhysRevLett.116.031301} {\bibfield  {journal} {\bibinfo
  {journal} {Phys. Rev. Lett.}\ }\textbf {\bibinfo {volume} {116}},\ \bibinfo
  {pages} {031301} (\bibinfo {year} {2016})},\ \Eprint
  {http://arxiv.org/abs/1507.04744} {arXiv:1507.04744 [astro-ph.CO]}
  \BibitemShut {NoStop}%
\bibitem [{\citenamefont {Powell}\ \emph {et~al.}(2017)\citenamefont {Powell},
  \citenamefont {Laha}, \citenamefont {Ng},\ and\ \citenamefont
  {Abel}}]{Powell:2016zbo}%
  \BibitemOpen
  \bibfield  {author} {\bibinfo {author} {\bibfnamefont {D.}~\bibnamefont
  {Powell}}, \bibinfo {author} {\bibfnamefont {R.}~\bibnamefont {Laha}},
  \bibinfo {author} {\bibfnamefont {K.~C.~Y.}\ \bibnamefont {Ng}}, \ and\
  \bibinfo {author} {\bibfnamefont {T.}~\bibnamefont {Abel}},\ }\href {\doibase
  10.1103/PhysRevD.95.063012} {\bibfield  {journal} {\bibinfo  {journal} {Phys.
  Rev. D}\ }\textbf {\bibinfo {volume} {95}},\ \bibinfo {pages} {063012}
  (\bibinfo {year} {2017})},\ \Eprint {http://arxiv.org/abs/1611.02714}
  {arXiv:1611.02714 [astro-ph.CO]} \BibitemShut {NoStop}%
\bibitem [{\citenamefont {Zhong}\ \emph {et~al.}(2020)\citenamefont {Zhong},
  \citenamefont {Valli},\ and\ \citenamefont {Abazajian}}]{Zhong:2020wre}%
  \BibitemOpen
  \bibfield  {author} {\bibinfo {author} {\bibfnamefont {D.}~\bibnamefont
  {Zhong}}, \bibinfo {author} {\bibfnamefont {M.}~\bibnamefont {Valli}}, \ and\
  \bibinfo {author} {\bibfnamefont {K.~N.}\ \bibnamefont {Abazajian}},\ }\href
  {\doibase 10.1103/PhysRevD.102.083008} {\bibfield  {journal} {\bibinfo
  {journal} {Phys. Rev. D}\ }\textbf {\bibinfo {volume} {102}},\ \bibinfo
  {pages} {083008} (\bibinfo {year} {2020})},\ \Eprint
  {http://arxiv.org/abs/2003.00148} {arXiv:2003.00148 [astro-ph.HE]}
  \BibitemShut {NoStop}%
\end{thebibliography}%

\end{document}